\documentclass[useAMS,usenatbib]{mnras}
\usepackage{graphicx,natbib,amssymb,amsmath,stmaryrd,makecell}
\usepackage{txfonts}
\usepackage{xcolor}
\usepackage{hyperref}
\usepackage{xspace}
\usepackage{caption}
\usepackage{lipsum}
\usepackage{soul}

\newcommand{\id}{{\,\rm d}}

\newcommand{\beq}{\begin{equation}}   %

\newcommand{\eeq}{\end{equation}}   %

\newcommand{\beqa}{\begin{eqnarray}}   %

\newcommand{\eeqa}{\end{eqnarray}}   %

\newcommand{\beal}{\begin{align}}

\newcommand{\enal}{\end{align}}

\newcommand{\bspl}{\begin{split}}

\newcommand{\espl}{\end{split}}

\newcommand{\bsub}{\begin{subequations}}

\newcommand{\esub}{\end{subequations}}

\newcommand{\bmulti}{\begin{multline}}   %

\newcommand{\beqm}{\begin{mathletters}}   %

\newcommand{\eeqm}{\end{mathletters}}   %

\newcommand{\COBEF}{{\it COBE/FIRAS}\xspace}

\author[Cyr, Acharya and Chluba]
{Bryce Cyr$^1$\thanks{E-mail:bryce.cyr@manchester.ac.uk}, Sandeep Kumar Acharya$^2$ and Jens Chluba$^1$\\
$^1$Jodrell Bank Centre for Astrophysics, School of Physics and Astronomy, The University of Manchester, Manchester M13 9PL, U.K.\\
$^2$Astrophysics Research Center of the Open University, The Open University of Israel, Ra'anana, Israel}
\date{\vspace{-0mm}Accepted XXX. Received YYY; in original form ZZZ}

\title[Soft photon heating]
{Soft Photon Heating: A Semi-Analytic Framework and Applications to $21$cm Cosmology}

\begin{document}
\maketitle
\begin{abstract}
    The presence of an abundant population of low frequency photons at high redshifts (such as a radio background) can source leading order effects on the evolution of the matter and spin temperatures through rapid free-free absorptions. This effect, known as soft photon heating, can have a dramatic impact on the differential brightness temperature, $\Delta T_{\rm b}$, a central observable in $21$cm cosmology. Here, we introduce a semi-analytic framework to describe the dynamics of soft photon heating, providing a simplified set of evolution equations and a useful numerical scheme which can be used to study this generic effect. We also perform quasi-instantaneous and continuous soft photon injections to elucidate the different regimes in which soft photon heating is expected to impart a significant contribution to the global $21$cm signal and its fluctuations. We find that soft photon backgrounds produced after recombination with spectral index $\gamma > 3.0$ undergo significant free-free absorption, and therefore this heating effect cannot be neglected. The effect becomes stronger with steeper spectral index, and in some cases the injection of a synchrotron-like spectrum ($\gamma = 3.6$) can suppress the amplitude of $\Delta T_{\rm b}$ relative to the standard model prediction, making the global $21$cm signal even more difficult to detect in these scenarios. \\
\end{abstract}

\section{\label{sec:level1}Introduction}
Heating mechanisms in cosmology can take many different forms, and a rigorous study by the community has rewarded us with a well-established understanding of the thermal history of the universe. At very early times ($10^3 \lesssim z\lesssim 10^{6}$), sources of energy injection are well-known to produce distortions to the frequency spectrum of the Cosmic Microwave Background (CMB) \citep{Zeldovich1969, Sunyaev1970, Illarionov1974, Danese1982, Burigana1991, Hu1993, Chluba2011therm}. These CMB spectral distortions are often characterized to be of $\mu$- or $y$-type, and measurements made by the \COBEF satellite in the 90s has led to constraints on a  variety of cosmological phenomena that remain competitive even to this day. 

At lower redshifts, around $z \simeq 10^3$, excess heating can manifest itself as a delay to the recombination process. This can have adverse effects on CMB anisotropy measurements [e.g. as made by the \citet{Planck2018params}], as well as on the spectrum of cosmological recombination radiation, allowing stringent constraints to be placed on models which disrupt this process \citep{Slatyer2009, Chluba2010a, Finkbeiner2011, Galli2013, Diamanti2013, Carr2020}. 

Another milestone in our thermal history, and the focus of this study, is the epoch of cosmic dawn and reionization [$z\simeq \mathcal{O}(10)$], where observations of redshifted $21$cm photon emission from spin-flip transitions can be studied to infer the distribution of neutral hydrogen. Reionization has seen a flurry of experimental and theoretical progress over the last decade, allowing it to develop rapidly into a mature subfield of cosmology \citep{Mellema2012, Jacobs2014, HERA2016, Cohen2017, Edges2018,Saras2022}. The central observable in this field is known as the differential brightness temperature, $\Delta T_{\rm b}$, which measures the contrast between the the temperature of the radiation bath at $21$cm frequencies ($T_{\rm R}$), and an object known as the spin temperature, $T_{\rm S}$ which measures the relative occupation of the spin triplet and singlet states of neutral hydrogen (see \citet{Furlanetto2006, Pritchard2011, Barkana2016} for comprehensive reviews).

This differential brightness temperature can be probed in a sky-averaged sense (a global signal, analogous to determining the CMB temperature), as well as by studying its spatial fluctuations (similar to the measurement of CMB temperature anisotropies). An initial detection of the global signal was reported by the EDGES collaboration \citep{Edges2018} which indicated that the $\Delta T_{\rm b}$ absorption trough was roughly twice as deep as was to be expected when considering standard astrophysical and cosmological evolution. Naturally, this spurred much development on the theory side to elucidate how one could either lower $T_{\rm S}$ \citep{Barkana2018, Munoz2018, Kovetz2018} or increase $T_{\rm R}$ \citep{Feng2018, Ewall-Wice2018, Brandenberger2019, Fialkov2019}. Recently, the SARAS-3 experiment \citep{Singh2021} reported results from their radiometer experiment, ultimately refuting the claimed detection made by the EDGES collaboration. While the observational study of the global signal has undergone some tumultuous times, fluctuation experiments are well under way, with the Hydrogen Epoch of Reionization Array (HERA, \citet{HERA2016}), recently providing some truly impressive limits on the power spectrum \citep{HERA2022b}.

The surface of last scattering presents us with a snapshot of the primordial perturbations, from which we have gleamed some of the most stringent constraints on parameters in the standard cosmological model ($\Lambda$CDM). In contrast, observations of the 21cm signal can be made at a variety of redshifts, providing us with a sensitive tomographic probe of the evolution of neutral hydrogen across cosmic time. Unlike the CMB, whose perturbations are set mainly by linear physics, a reliable computation of $\Delta T_{\rm b}$ requires input from complex astrophysical modeling, often occurring in the non-linear regimes of structure formation. Thus, in order to fully take advantage of the impending mountain of data promised to us by 21cm experiments, it is imperative that we understand the various ways in which heating, both of the photon background and of the hydrogen gas, can occur.

In this work, we present a semi-analytic formalism to determine the heating rate of the hydrogen gas at all times after recombination (including during the epochs of cosmic dawn and reionization) in the presence of a radio (soft photon) background. This effect was assumed small and neglected in previous studies on the impacts of extra radio backgrounds in 21cm cosmology (e.g. in \cite{Feng2018, Ewall-Wice2018, Fialkov2019}). Instead, we find that there are many scenarios in which this heating via soft photon backgrounds produces a leading order effect on the determination of $\Delta T_{\rm b}$, and can therefore not be neglected.

This mechanism, which we call \textit{soft photon heating}, can be roughly summarized as follows: the presence of a sufficiently steep soft photon background in place before cosmic dawn ($z \gtrsim 20$) is capable of greatly lowering the contrast between $T_{\rm R}$ and $T_{\rm S}$, in turn reducing the amplitude of $\Delta T_{\rm b}$. The physical effect neglected in previous studies is that of free-free absorption, which becomes extremely efficient at low frequencies and can cause a significant boost to the gas temperature. 

Recently \citep{Acharya2023SPH}, we studied this effect using the (fully numeric) thermalization code, \texttt{CosmoTherm}\footnote{\url{www.chluba.de/CosmoTherm}}. This tool, originally developed to perform detailed calculations of CMB spectral distortions \citep{Chluba2011therm}, has been augmented over the past years to study a more versatile range of cosmological effects, including the global 21cm signal for both standard and exotic phenomena \citep{ADC2022}. Using this machinery, we produced constraints from a variety of observational probes using \texttt{CosmoTherm} for decaying dark matter scenarios \citep{Acharya2023SPH}, superconducting cosmic strings \citep{Cyr2023CSS,Cyr2023a}, and more generic broad-spectrum photon injections \citep{Acharya2023b}.

Here, we provide a simplified set of coupled integro-differential equations which must be solved to properly incorporate the effects of soft photon heating. This formalism has been implemented in {\tt CosmoTherm} for quite some time as a way to simplify the computation at $z\lesssim 10^3$ \citep{Chluba2015GreensII}. Here we further expand upon these approximations and illustrate some important new effects related to soft photon heating. In addition, we provide an efficient numerical scheme for solving the photon evolution equation, which can readily be implemented into standard $21$cm codes \citep{Mesinger2010, Fialkov2015, Munoz21cm2023}. After developing the formalism, we inject soft photon backgrounds both quasi-instantaneously and continuously, showcasing the non-trivial evolution of the spin, matter, and radiation temperatures. Using these models, we study changes to $\Delta T_{\rm b}$ when soft photon heating is switched on and off. We also provide a simple Jupyter Notebook which one can use to qualitatively study this heating in the presence of quasi-instantaneous injections\footnote{\url{https://github.com/Bryce-Cyr/Soft_Photon_Heating}}.

The rest of the paper is organized as follows. In Section \ref{sec:level2}, we briefly review the various heating mechanisms relevant to the computation of $\Delta T_{\rm b}$ finishing with a more qualitative discussion on the mechanism and implications of soft photon heating. We proceed in Sec. \ref{sec:level3} with a derivation of the system of coupled integro-differential equations which describes the effect, providing a useful numerical scheme. Section \ref{sec:standard_ff} describes the relative (un)importance of soft photon heating when the CMB is the unique source of low frequency photons. Next, in Section \ref{sec:level4}, we perform quasi-instantaneous soft photon injections, with different injection redshifts and spectral indices, studying the evolution of various quantities of interest to $21$cm cosmology. This allows us to elucidate the conditions under which soft photon heating cannot be ignored. Section \ref{sec:cont_inj} repeats this analysis for continuous injection scenarios, motivated by decaying dark matter and cosmic string models. The $21$cm power spectrum is discussed in Section \ref{sec:level6}, and we conclude in Section \ref{sec:level7}.

\section{Heating mechanisms in 21cm cosmology}
\label{sec:level2}
The differential brightness temperature is central to $21$cm observations, and its amplitude is determined by the contrast between the radiation and spin temperatures at a given redshift
%
\begin{align}
    \Delta T_{\rm b} &= \frac{T_{\rm S} - T_{\rm R}}{1+z} \left( 1 - {\rm e}^{-\tau_{21}}\right), 
\end{align}
%
where $\tau_{21}$ is the 21cm optical depth. This can be approximated as \citep{BL2005,MFC2011},
%
\begin{align} \label{eq:dTb}
    \Delta T_b \approx 27x_{\rm H}(1+\delta)\left(\frac{1+z}{10}\frac{0.15}{\Omega_m h^2}\right)^{1/2}\left(\frac{\Omega_b h^2}{0.023}\right) \nonumber \\
    \left(\frac{H}{{\rm dv_r/dr}+H}\right) \left(1-\frac{T_{\rm R}}{T_{\rm S}}\right)\hspace{0.2cm} {\rm mK},
\end{align}
%
where $x_{\rm H}$ is the neutral hydrogen fraction, $\delta$ is the fractional overdensity of the baryons, and ${\rm dv_r}/{\rm dr}$ is the velocity gradient along the line of sight. Spatial fluctuations of $\Delta T_{\rm b}$ exist and represent a valuable area of study, but for the purpose of this illustration we will focus on the case of a global signal, returning to the fluctuations in Section \ref{sec:level6}. The amplitude of the differential brightness temperature is thus proportional to $\Delta T_{\rm b} \propto x_{\rm H}(1-T_{\rm R}/T_{\rm S})$. The radiation temperature is set by the intensity of the photon background at $21$cm frequencies,
%
\begin{align} \label{eq:Trad}
    T_{\rm R}(z) = \left.\frac{c^2}{2 k_{\rm b}} \frac{I_{\nu}(z)}{\nu^2}\right|_{\nu = \nu_{\rm 21}},
\end{align}
%
where $c$ and $k_{\rm b}$ are the speed of light and Boltzmann constant respectively, while $\nu_{\rm 21}$ is the rest frame 21cm frequency (1.42 GHz). In standard scenarios the intensity spectrum is simply set by the CMB, while modifications are necessary when one introduces additional soft photon (radio) backgrounds. 

The spin temperature counts the relative number of hydrogen atoms in the spin triplet state vs the singlet, and is often expressed as the equilibrium balance of processes which can induce spin-flip transitions
%
\begin{align} \label{eq:Tspin}
    T_{\rm S}^{-1} = \frac{x_{\rm R} T_{\rm R}^{-1} + x_{\rm c} T_{\rm M}^{-1}+x_{\alpha} T_{\alpha}^{-1}}{x_{\rm R} + x_{\rm c} + x_{\alpha}}.
\end{align}
%
Here, $x_{\rm R}$, $x_{\rm c}$ and $x_\alpha$ are the radiative, collisional and Wouthuysen-Field coupling coefficients respectively, $T_{\rm M}$ is the matter temperature, and $T_{\alpha}$ is the colour temperature of the Ly-$\alpha$ radiation field (see \cite{Pritchard2011} for a detailed review, and \citet{Venu2018} for a discussion on the precise form of $x_{\rm R}$). The addition of extra radio backgrounds therefore directly propagates in this way to $T_{\rm S}$.

At high redshifts ($z \gtrsim 150$), Compton scattering between the electrons and background photons drive the temperatures of these two sectors to be roughly equal, $T_{\rm M} \simeq T_{\rm CMB}$, strongly suppressing any possible deviation of $\Delta T_{\rm b}$ from 0. At lower redshifts, Compton scattering becomes inefficient, and the spin temperature (which is initially tightly coupled to $T_{\rm M}$) deviates from $T_{\rm R}$, producing the well known absorption trough feature in $\Delta T_{\rm b}$. At even later times, the first stars begin to form, sourcing an abundant population of X-rays. This hard photon injection provides an efficient heat source for the hydrogen gas, driving the spin temperature back to $T_{\rm R}$ \citep{Venkatesan2001, Pritchard2006, Zaroubi2006}, marking an end to the epoch known as cosmic dawn. 

The mean free path of sufficiently strong X-rays ($E \gtrsim 1$ keV) can be cosmological, implying that they are capable of depositing their energy at distances far from their source. X-ray binaries produced from stellar remnants appear to be a plausible source of heating around the time of cosmic dawn, and the spectrum of X-rays produced by these binaries plays a major role in the exact timing of of the $\Delta T_{\rm b}$ absorption trough, as discussed in \citet{Fialkov2014, Pacucci2014}. 

Additional heating mechanisms have also been studied. For example, a number of authors have considered cosmic rays as another source of early heat deposition into the intergalactic medium (IGM). In a variety of works \citep{Sazonov2015,Leite2017,Yokoyama2023}, it was shown that these cosmic rays can provide a significant increase to the gas temperature at high enough redshifts to impact the observables relevant to cosmic dawn and reionization. Overall, one expects a reduction in the amplitude of $\Delta T_{\rm b}$ due to the increased gas temperature, as noted in \citet{Jana2018}.

More recently, \citet{Gessey-Jones2023} performed the first detailed simulations of this setup and found that cosmic rays deposit most of their energy nearer to the source than X-rays. In principle, this allows for component separation between X-ray and cosmic ray heating by leveraging 21cm power spectrum data and tomographic maps. In practice, however, the cosmic ray scenarios still possess a large number of theoretical uncertainties which makes performing detailed predictions difficult. In particular, both \citet{Leite2017} and \citet{Gessey-Jones2023} found that modest variations in the assumed spectral index of cosmic rays caused large deviations in the IGM temperature at redshifts as low as $z \simeq 8$.

Around the time of the EDGES measurement \citep{Edges2018}, it was thought that the presence of an extra radio background with synchrotron-like spectral index could provide large enhancements to $\Delta T_{\rm b}$ \citep{Feng2018, Ewall-Wice2018, Brandenberger2019, Fialkov2019} through an increase in $T_{\rm R}$. What was not realized at the time is that the abundance of low frequency photons present in such a background can cause significant heating of the IGM through efficient free-free absorption. Heating through this effect is typically subdominant in standard scenarios (e.g. when the CMB sets the amplitude of the low frequency photon background), but this changes dramatically when a significant population of soft photons is introduced. Thus, $T_{\rm R}$, $T_{\rm M}$, and $T_{\rm S}$ are not as decoupled as originally thought, and must be evolved together to properly account for this effect. It is this joint evolution which we call \textit{soft photon heating}.

In previous work, we studied this in a fully numerical setup, and found that for a sufficiently energetic synchrotron background, the ratio of $T_{\rm R}/T_{\rm S}$ was greatly suppressed relative to the case when soft photon heating was neglected \citep{Acharya2023SPH}. When examining a toy model of the synchrotron spectrum which closely matched the ARCADE-2 \citep{Fixsen2011excess} and LWA-1 \citep{DT2018} radio background data, this effect caused a dramatic suppression to the amplitude of $\Delta T_{\rm b}$. 

In the following, we revisit this result and provide a semi-analytic scheme to determine the non-trivial evolution of the matter temperature in the presence of a soft photon background. After developing the formalism, we compute the spin temperature evolution for a number of models, utilizing a simple reionization module with a standard population of Ly-$\alpha$ fluxes. Details of this module can be found in \citet{ADC2022}. Moreover, in order to disentangle soft photon heating from other mechanisms, we neglect additional modeling extensions of the early sources, such as the inclusion of cosmic ray heating.

\section{Soft photon heating: a semi-analytic approach}
\label{sec:level3}

The transfer of energy and particle number between photons and electrons is a setup which has been carefully studied in the context of CMB spectral distortions. An intimate knowledge of this thermalization procedure is necessary to understand which types of effects remain frozen in on the photon spectrum, providing observational signatures one can search for. A deficit of low-energy photons due to free-free absorption is one such signature, thus it is perhaps unsurprising that our starting point comes from the spectral distortion literature \citep{Chluba2015GreensII}.

A word on preliminaries: when referring to photon frequencies, we often make use of the dimensionless form $x = h\nu/k_{\rm b}T_{\rm CMB}$, which has the convenient property that it is redshift invariant, due to the identical scalings of $\nu$ and $T_{\rm CMB}$ with $z$. The peak of a blackbody spectrum has $x_{\rm peak} \approx 2.82$, while data on the radio synchrotron background \citep{Fixsen2011excess, DT2018} sits at roughly $10^{-4} \lesssim x \lesssim 10^{-1}$. We will often work in terms of occupation numbers instead of intensities, which are related through
%
\begin{subequations}
\begin{align}
    I_{\rm bb}(x, T_{\rm CMB}) &= \frac{2 (k_{\rm b}T_{\rm CMB})^3}{(h c)^2} x^3 n_{\rm bb}(x), \label{eq:Intensity-occupation}\\
    n_{\rm bb}(x) &= \frac{1}{{\rm e}^x - 1}.
\end{align}
\end{subequations}
%
For a general (non-thermal) intensity spectrum, the same relation holds, such that $I \propto x^3 n(x)$ where $n(x)$ is a non-blackbody occupation number. Keep in mind that $T_{\rm CMB} =T_{\rm CMB, 0}(1+z)$, which causes the intensity spectrum to retain a redshift dependence.

Additionally, we define an analogous form for photons frequencies at the electron temperature $x_{\rm e} = x \, T_{\rm CMB} / T_{\rm M}$ which will be computationally convenient later on. This quantity does inherit a redshift dependence once the CMB and matter temperatures decouple at $z \lesssim 150$. As a caution, this should not be confused with $X_{\rm e}$, which we will take to define the free electron fraction. Finally, we will make use of the Thomson scattering optical depth, $\id \tau = \sigma_{\rm T} N_{\rm e} c \, \id t$ as a proxy for time evolution. Here, $\sigma_{\rm T}$ is the Thomson scattering cross section and $N_{\rm e}$ the free electron number density.

\subsection{Free-free absorption at low frequencies}
We begin our formulation of the photon evolution equation at times when Compton scattering is quite inefficient at redistributing energy throughout the photon spectrum (i.e. $z \lesssim 1500$). At these redshifts, the evolution of the photon occupation number is primarily influenced by free-free emission and absorption, and is given by \citep[e.g., see][]{Chluba2015GreensII}
%
\begin{equation} \label{eq:starting-point}
    \frac{\id n}{\id \tau} \simeq  \frac{\Lambda_{\rm BR}(\tau,x_{\rm e}) \,{\rm e}^{-x_{\rm e}}}{x_{\rm e}^3}\left[ 1 - n({\rm e}^{x_{\rm e}} - 1)\right] + S(\tau,x).
\end{equation}
%
Here, $\Lambda_{\rm BR}(\tau, x_{\rm e})$ is the (dimensionless) Bremsstrahlung emissivity coefficient and $S(\tau,x)$ is a source term that we will elaborate upon briefly. A derivation of this equation from first principles, as well as the details of $\Lambda_{\rm BR}(\tau,x_{\rm e})$ can be found in Appendix \ref{sec:levelA1}. The inefficacy of Compton scattering at $z \lesssim 1500$ allows us to neglect mode couplings between different photon frequencies, greatly simplifying the thermalization problem \citep{Chluba2015GreensII}.

We will be most interested in low-frequency departures from a blackbody, so it is useful to consider only the part that deviates from a blackbody, $\Delta n = n - n_{\rm bb}$, the so-called spectral distortion. The evolution of this spectral distortion is
%
\begin{equation} \label{eq:dn-evolution}
    \frac{\id \Delta n}{\id \tau} \simeq  -\frac{\Lambda_{\rm BR} (1 - {\rm e}^{-x_{\rm e}})}{x_{\rm e}^3} \Delta n + \Delta S(\tau, x).
\end{equation}
%
The first term encodes the effects of free-free absorption and emission in the presence of a distortion, governing the efficiency of soft photon heating. We have absorbed the terms not dependent on the distortion into a newly defined source term, $\Delta S$,
%
\begin{align} \label{eq:source-term}
    \Delta S(\tau,x) &= S_{\rm inj}(\tau,x) + \frac{k_{\rm b}}{m_{\rm e} c^2}(T_{\rm M} - T_{\rm CMB}) Y(x) + S_{\rm ff,bb}(\tau,x),\\
    Y(x) &= \frac{x {\rm e}^x}{({\rm e}^x - 1)^2}\left[ x \frac{{\rm e}^x + 1}{{\rm e}^x - 1}-4 \right]. \nonumber
\end{align}
%
The first term is simply the rate of occupation number injection for whichever low-frequency background we choose to study, $S_{\rm inj} = \id n_{\rm inj}/\id \tau$. The electron population cools faster than the photon background, but at $z \gtrsim 150$ residual Compton scatterings are able to keep $T_{\rm M} \simeq T_{\rm CMB}$. This energy extraction from the CMB photons generically sources a distortion, which is enforced by the second term in Eq.~\eqref{eq:source-term}, with $Y(x)$ being the spectral parameterization of the $y$-distortion. A full derivation of this effect is also presented in Appendix \ref{sec:levelA1}. Finally, the third term accounts for free-free emission and absorption off of the blackbody itself. Its form is given by
%
\begin{align}
    S_{\rm ff,bb}(\tau,x) = \frac{\Lambda_{\rm BR} (1-{\rm e}^{-x_{\rm e}})}{x_{\rm e}^3} \left[ \frac{1}{{\rm e}^{x_{\rm e}}-1}-\frac{1}{{\rm e}^{x}-1}\right].
\end{align}
%
From this expression it is evident that when the electron temperature is higher than the CMB temperature, free-free effects lead to the emission of photons, while photons are absorbed when $T_{\rm M}<T_{\rm CMB}$. Energy transfer between these blackbody photons and the electrons is typically quite small, as we will see in Sec. \ref{sec:standard_ff}. Now, let us define the free-free optical depth 
%
\begin{align}
    \tau_{\rm ff} = \int_0^{\tau} \frac{\Lambda_{\rm BR}(\tau',x_{\rm e})(1-{\rm e}^{-x_{\rm e}})}{x_{\rm e}^3} \id \tau'.
\end{align}
%
By performing a change of variable to $\tau_{\rm ff}$ in Eq.~\eqref{eq:dn-evolution}, the evolution equation is in a form that can easily be solved using integrating factors. The result is \citep[e.g., similar to][]{Chluba2015GreensII}
%
\begin{align} \label{eq:dn-sol}
    \Delta n(x,\tau_{\rm ff,obs}) = \Delta n(x,0) \,{\rm e}^{-\tau_{\rm ff,obs}} + {\rm e}^{-\tau_{\rm ff,obs}} \int_0^{\tau_{\rm ff,obs}} \id \tau_{\rm ff}' \, \Delta \tilde{S} \,{\rm e}^{\tau_{\rm ff}'}.
\end{align}
%
In this expression, $\tau_{\rm ff,obs} = \tau_{\rm ff}(z_{\rm obs})$ is the free-free optical depth at some target redshift, and $\tau_{\rm ff,ini} = 0$. For convenience, the source term has also been slightly redefined
%
\begin{align}
    \Delta \tilde{S} = \frac{x_{\rm e}^3}{\Lambda_{\rm BR}(1-{\rm e}^{-x_{\rm e}})} \Delta S. \nonumber
\end{align}
%
Equation~\eqref{eq:dn-sol} illustrates how a spectral distortion injected at some initial time [$\tau_{\rm ff}(z_{\rm ini}) = 0$] is attenuated as it propagates to later times. Distortions which are sourced post-recombination with $x \gtrsim 10^{-4}$ do not exhibit a large amount of free-free absorption \citep{Chluba2015GreensII, Bolliet2020PI}, as we will also show below. This situation changes dramatically at lower frequencies due to the sharp increase in the absorption cross section. For numerical schemes, Eq.~\eqref{eq:dn-sol} can be recast in a particularly useful way
%
\begin{align} \label{eq:dn-numeric}
    \Delta n(x,\tau_{\rm ff, i+1}) \approx \Delta n(x, \tau_{\rm ff, i}) \, {\rm e}^{-\Delta \tau_{\rm ff}} + \Delta \tilde{S} \left( 1 - {\rm e}^{-\Delta \tau_{\rm ff}}\right),
\end{align}
%
where $\Delta \tau_{\rm ff} = \tau_{\rm ff,i+1} - \tau_{\rm ff,i}$ is the size of a given timestep. The condition for the validity of this iterative process is that the source term $\Delta \tilde{S}$ must be roughly constant over a given timestep, where $\Delta \tau_{\rm ff}$ need not be small. We have found that redshift step sizes of $\Delta z \lesssim 0.5$ offer a good balance of convergence to the full result (computed using \texttt{CosmoTherm}), and fast runtimes\footnote{The reader is encouraged to check this themselves by running the provided Jupyter Notebook.}.

\subsection{Evolution of the matter temperature}
The next step is to understand the response of the matter temperature to (potentially sizeable) free-free absorptions. In the absence of interactions with the photons, adiabatic cooling of the matter induces the well known scaling $T_{\rm M} \propto (1+z)^2$. At early enough times ($z \gtrsim 150$), Compton scattering off the background provides an efficient heat source for the gas, enforcing $T_{\rm M} \simeq T_{\rm CMB}$. This only requires the presence of the usual CMB, and produces a weak ({\it negative}) $y$-distortion to the blackbody spectrum \citep{Chluba2005, Chluba2011therm}. In the presence of additional soft photon backgrounds, supplemental gas heating can occur and the matter temperature evolves as
%
\begin{align} 
\label{eq:T_M}
    \frac{{\rm d}T_M}{{\rm d}z}=\frac{2T_M}{1+z}+\frac{X_{\rm e}}{1+X_{\rm e}+f_{\rm He}}\frac{8\sigma_{\rm T}\rho_{\rm CMB}}{3m_{\rm e} c}\frac{T_{\rm M}-T_{\rm CMB}}{(1+z)H(z)}+\frac{{\rm d}T_{\rm ff}}{{\rm d}z}.
\end{align}
%
In this expression, $X_{\rm e} = N_{\rm e}/N_{\rm H}$ is the free electron fraction, $f_{\rm He}\simeq 0.08$ is the helium fraction by number of nuclei, $N_{\rm H}$ is the number density of hydrogen, $\rho_{\rm CMB}$ is the (physical) energy density of the CMB at temperature $T_{\rm CMB}$, and $m_{\rm e}$ is the electron mass. Note that a positive contribution, $\id T_{\rm ff}/\id z > 0$ corresponds to a cooling of the gas temperature, while a negative contribution sources heating. The net effect caused by the reprocessing of a soft photon background through free-free is 
%
\begin{align} \label{eq:T-ff}
    \frac{\id T_{\rm ff}}{\id z} = -\frac{1}{(3/2) k_{\rm b} N_{\rm H}(1+X_{\rm e} + f_{\rm He})} \frac{1}{a^4}&\frac{\id a^4 \Delta \rho_{\rm ff}}{\id z},
\end{align}
%
where the denominator represents the heat capacity of the medium, and $\id (a^4 \Delta \rho_{\rm ff})/ \id z$ is the comoving change of the photon energy density by the free-free part of the photon evolution equation. We provide the derivation of the Compton heating term, as well as an explicit form for this free-free heating term in Appendix \ref{sec:levelB1}. For the latter, the final result is
\begin{align} \label{eq:ff-inj}
    &\frac{1}{a^4}\frac{\id a^4 \Delta \rho_{\rm ff}}{\id z} = -\frac{\rho_{\rm CMB}}{\pi^4/15} \left( \frac{T_{\rm M}}{T_{\rm CMB}}\right)^3 \frac{\sigma_{\rm T} N_{\rm e} c}{H(1+z)} \\
    &\times \int_0^{\infty} \id x \, \Lambda_{\rm BR}(\tau, x_{\rm e}) \left( 1 - {\rm e}^{-x_{\rm e}}\right) \left[\frac{1}{{\rm e}^{x_{\rm e}}-1} - \frac{1}{{\rm e}^{x}-1} -\Delta n(x) \right]. \nonumber
\end{align}
The injection of a soft photon background is achieved through a large and positive contribution to $\Delta n(x)$ at low frequencies. After injection, these photons are reprocessed and absorbed by the medium, yielding important contributions to the heating term, $\id T_{\rm ff}/\id z$.

\subsection{Evolution of the electron fraction}
The final piece of this puzzle requires us to understand the evolution of the free-electron fraction in the presence of a soft photon background. Due to the high optical depth for low-frequency photons prior to recombination, we are primarily concerned with soft photon backgrounds generated at $z \lesssim 1500$. At these redshifts, helium recombination has almost entirely completed \citep{Switzer2007I, Kholupenko2007, Jose2008}, so we choose to neglect any effects associated with it. The evolution of the free-electron fraction is determined by \citep{SeagerRecfast1999, Seager2000}
%
\begin{align}
    \frac{{\rm d}X_{\rm e}}{{\rm d}z}=\frac{\big[\alpha_{\rm H} X_{\rm e}^2 N_{\rm H}-\beta_{\rm H}(1-X_{\rm e})B_{\rm H}\big]C_{\rm H}}{(1+z)H(z)},
    \label{Eq:recombination}
\end{align}
%
where $\alpha_{\rm H}$ and $\beta_{\rm H}$ are the Case B recombination and photoionization rates respectively, $B_{\rm H} = {\rm exp}[-E_{\alpha}/k_{\rm b} T_{\rm CMB}]$ (here $E_{\alpha} = 10.2 \, {\rm eV}$ is the energy of a Ly-$\alpha$ photon), and the Peebles factor is given by
%
\begin{align}
    C_{\rm H}=\frac{1+K_{\rm H} A_{\rm 2s1s,H}N_{\rm H}(1-X_{\rm e})}{1+K_{\rm H}(A_{\rm 2s1s,H}+\beta_{\rm H})N_{\rm H}(1-X_{\rm e})}.
\end{align}
%
In this expression, $A_{\rm 2s1s,H} = 8.22458 \, {\rm s}^{-1}$ is the decay rate of the 2s level to the ground state, and $K_{\rm H} = \lambda_{\alpha}^3/8\pi H(z)$ where $\lambda_{\alpha}$ is the wavelength of a Ly-$\alpha$ photon. Additionally, the photoionization rate is related to the recombination rate through
%
\begin{align}
    \beta_{\rm H} = \alpha_{\rm H}(T_{\rm CMB}) \left(\frac{m_{\rm e} k_{\rm b}T_{\rm CMB}}{2\pi \hbar^2}\right)^{3/2} \, {\rm e}^{- E_{\rm 2s}/k_{\rm b} T_{\rm CMB}}.
\end{align}
%
Here, $E_{\rm 2s} = 3.4 $ eV is the ionization energy from the 2s state. We make a special note that $\beta_{\rm H}$ should be evaluated at the radiation temperature, in contrast to $\alpha_{\rm H}$ which should be evaluated at $T_{\rm M}$. While this does not represent the full physics of the recombination and photonionization rates in a blackbody ambient radiation field \citep[e.g.,][]{Chluba2007, Grin2009}, it captures the leading order dependence more correctly \citep{Chluba2015PMF}. Thus, the recombination rate coefficient is sensitive to the matter temperature, such that higher $T_{\rm M}$ will lead to a higher frozen out fraction of $X_{\rm e}$.

In practice, we evolve the electron fraction using Eq.~\eqref{Eq:recombination} from a redshift of $z = 1500$ down to $z = 50$. At this point, we utilize the reionization module of \texttt{CosmoTherm} to further evolve $X_{\rm e}$, performing some simple modeling of the first sources of hard photons, important around the epoch of cosmic dawn. A full description of this module and its assumptions is given in \citet{ADC2022}.

Equation~\eqref{Eq:recombination} can be modified to also include the effects of collisional ionizations within the gas. This form of ionization typically only becomes important in extreme scenarios when $T_{\rm M} \gtrsim 10^4$~K \citep[e.g.,][]{Chluba2015PMF}. As we will see, this only happens in rather unphysical setups, and we neglect these corrections from our analytic description. We make further comments regarding this in Section \ref{sec:level4}.

Finally, source terms with significant emission above $E_{\rm inj} \gtrsim 10.2 \,{\rm eV}$, will induce direct excitations and ionizations to the gas. The usual partition in this case is that roughly $1/3$ of the energy injected above Ly-$\alpha$ frequencies will contribute to direct ionizations of the medium \citep{Chen2004}, with the rest of the energy contributing to excitations and heating. For most of the injection scenarios considered in this work, photons above this energy are never injected, allowing us to neglect direct ionizations and focus solely on soft photon heating. However, photon injections from decaying cosmic strings do possess a small fraction of these hard photons, which we discuss further in Sec. \ref{sec:cont_inj}.

Ultimately, the soft photon heating effect can be modeled by solving the coupled system of evolution equations defined by Eqs.~\eqref{eq:dn-evolution}, \eqref{eq:T_M}, and \eqref{Eq:recombination}. Without the presence of a source term ($S_{\rm inj} = 0$) our solutions reproduce the standard cosmological evolution, which includes free-free absorption and emission off of the CMB itself (this effect has always been included in {\tt CosmoTherm}). However, the inclusion of a soft photon background can dramatically modify the matter temperature, which in turn induces strong deviations to the expected $21$cm differential brightness temperature. In the cases that follow, we have checked that this simple setup faithfully captures the features of a more exact computation such as can be performed using \texttt{CosmoTherm}.

\vspace{-3mm}
\section{Matter heating rates and the CMB}
\label{sec:standard_ff}
\begin{figure}
\includegraphics[width=\columnwidth]{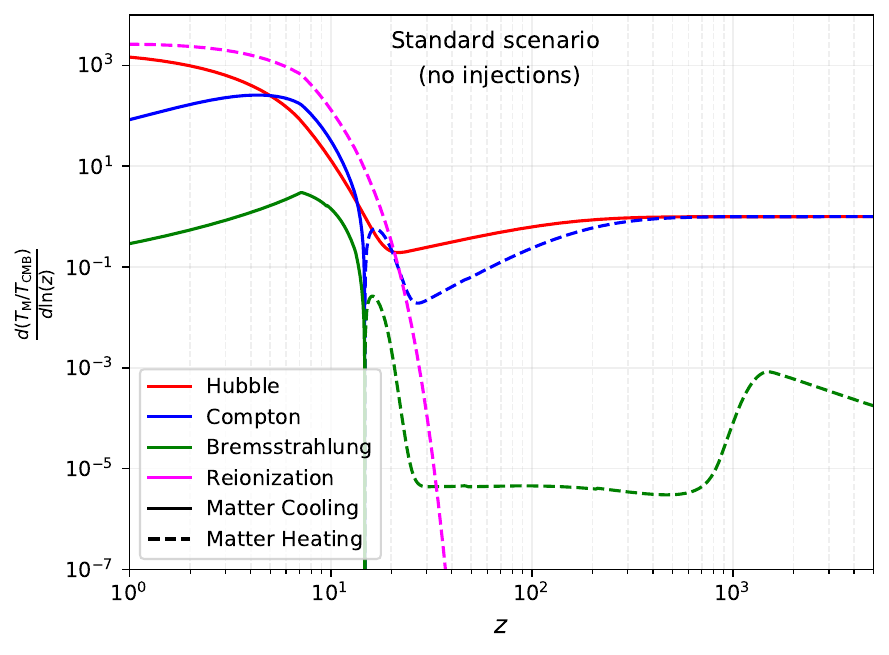}
\caption{Relative gas heating rates per logarithmic redshift interval in the presence of the CMB. The Hubble, Compton, and Bremsstrahlung curves correspond to the first, second, and third terms respectively in Eq.~\eqref{eq:T_M}. Solid lines indicate cooling of the ambient hydrogen gas, while the dashed correspond to heating.}
\label{fig:Vanilla-heating-rates}
\end{figure}

Before we consider exotic injections of a soft photon background, it is instructive to understand the relative importance of the individual heating rates in the standard scenario when only the CMB is present. The evolution of the matter temperature proceeds as usual following the prescription set out in Eq.~\eqref{eq:T_M}, specifically, through Hubble expansion, Compton scattering, and free-free absorption and emission. In Fig.~\ref{fig:Vanilla-heating-rates} we show the redshift dependence of each of these effects, where dashed lines indicate times when a particular interaction heats the gas, while solid lines highlight periods of gas cooling.
%
%

Focusing first on the Hubble and Compton contours, we see that at high redshifts ($z \gtrsim 150$), their respective cooling and heating rates almost completely compensate each other. It is during this period that residual Compton scatterings between the electrons and background photons source a well known (negative) $y$-type distortion, as the electrons continuously sap energy from the CMB. At $z \lesssim 150$, the Compton interaction rate drops rapidly and the matter temperature begins to evolve as $T_{\rm M} \propto (1+z)^2$ until reionization turns on at $z\simeq \mathcal{O}(10)$. At this point, the matter is rapidly heated (primarily by X-ray sources), leading to the small bump in the Compton heating rate around $z \simeq 20$ seen in Fig.~\ref{fig:Vanilla-heating-rates}. Shortly afterwards, the gas reaches temperatures $T_{\rm M} > T_{\rm CMB}$, and Compton cooling begins to occur. During this final phase, the electrons begin up-scattering the background photons, producing a small (positive) $y$-type distortion.
%
\begin{figure*}
\centering 
\includegraphics[width=\columnwidth]{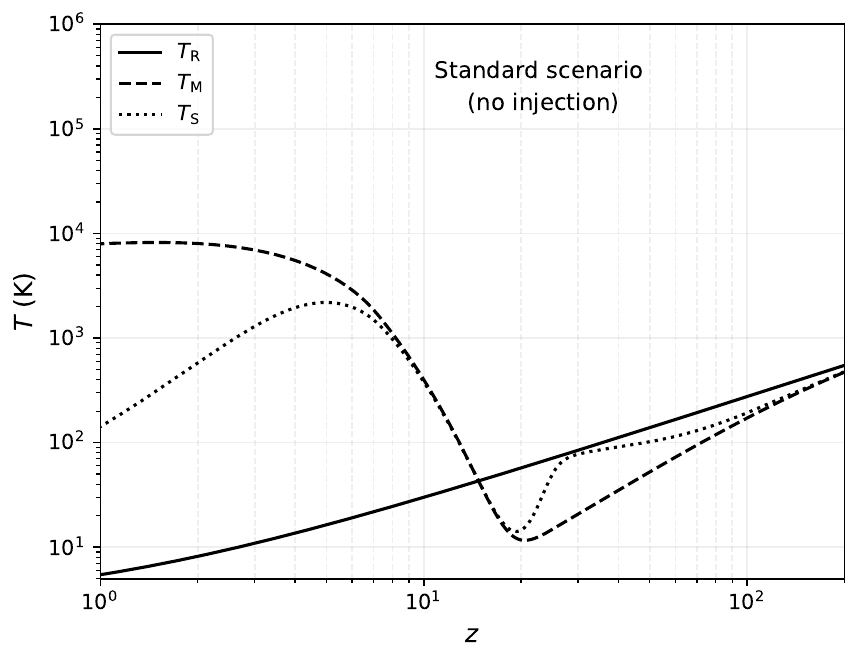}
\hspace{4mm}
\includegraphics[width=\columnwidth]{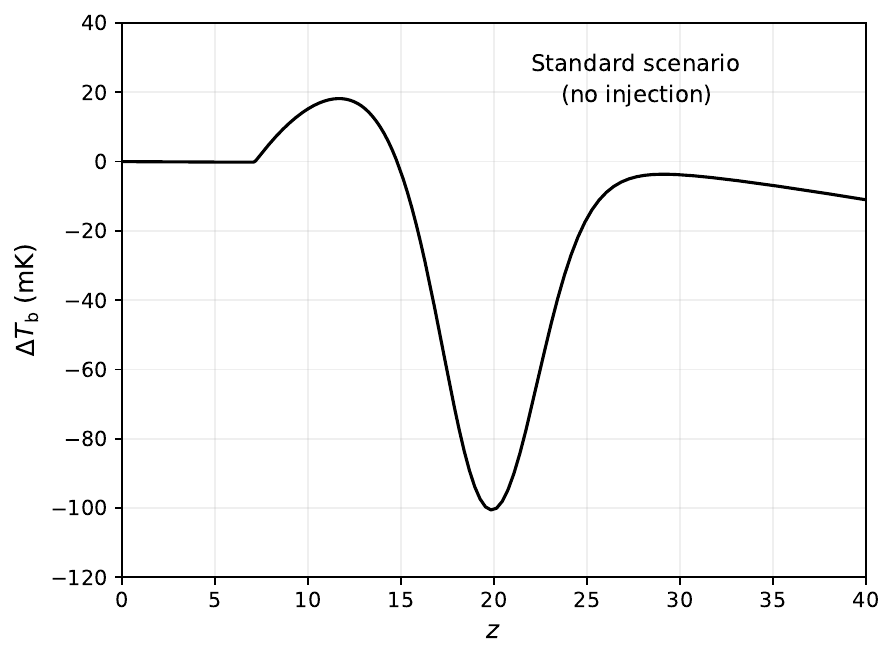}
\caption{Left: Redshift evolution of the radiation, matter, and spin temperatures in a standard $\Lambda$CDM scenario with no extra soft photon background. Right: the (global) differential brightness temperature. The amplitude of the absorption trough scales with $\Delta T_{\rm b} \propto x_{\rm H}(1-T_{\rm R}/T_{\rm S})$.}
\label{fig:dTb-Vanilla}
\end{figure*}

Turning our attention now to the free-free effects, the first thing to notice is that Bremsstrahlung typically plays a strongly subdominant role in the evolution of $T_{\rm M}$ without the addition of an extra soft photon background. At redshifts before recombination, the ionization fraction is $X_{\rm e} \simeq 1$ and low frequency photons are rapidly absorbed and emitted by the plasma. Note that free-free absorption will cause CMB photons (with temperature $T_{\rm abs} = T_{\rm CMB}$) to be absorbed by the gas, while the emission process will liberate a blackbody photon instead at the matter temperature ($T_{\rm emit} = T_{\rm M}$). Thus, the small deviation of matter and photon temperatures at early times leads to the weak (but steadily growing) soft photon heating at $z \gtrsim 1000$\footnote{An examination of Eq.~\eqref{eq:ff-inj} in the $\Delta n(x) = 0$ regime also indicates that heating occurs whenever $T_{\rm M} \lesssim T_{\rm CMB}$.}. After recombination, $X_{\rm e}$ drops rapidly which severely damps the free-free effects until reionization can rejuvenate the fraction of free electrons. At this point, a familiar situation occurs: soft photon heating boosts the gas temperature at an ever-increasing rate, until $T_{\rm M} > T_{\rm CMB}$, at which point free-free emission begins to provide a cooling contribution to $T_{\rm M}$.

The magenta contour represents the contribution to gas heating from the standard reionization module implemented through \texttt{CosmoTherm}. Thus far we have not accounted for any feedback from soft photon heating on the modeling assumptions of the first astrophysical sources, though this is a rich avenue for further study. As a result, in the following sections the heating curve from reionization will not be altered between the different injection scenarios. 

The left panel of Fig.~\ref{fig:dTb-Vanilla} showcases the familiar evolution of the radiation, spin, and matter temperatures in this case, while the right panel displays the predicted shape and amplitude of the differential brightness temperature, $\Delta T_{\rm b}$. At the moment, there is no undisputed detection of the global 21cm signal, though work by the EDGES and SARAS-3 collaborations have provided us with a loose lower limit of $\Delta T_{\rm b} \geq -500$ mK \citep{Edges2018,Saras2022}. Current literature \citep{Feng2018, Ewall-Wice2018, Fialkov2019} suggests that the presence of a radio background at cosmic dawn will strongly enhance the depth of the $\Delta T_{\rm b}$ absorption trough, making the signal much easier to detect. Stringent constraints have been placed on models which enhance the background temperature at $21$cm based on this lower limit \citep{Brandenberger2019, MRKD2021, Caputo2022b}. In the following two sections, we will show how induced soft photon heating from sufficiently steep and bright radio backgrounds can dramatically weaken these constraints, in some cases rendering the global $21$cm signal even more difficult to observe.

\section{Quasi-instantaneous injections}
\label{sec:level4}
In this work, we consider the impact of extra soft photon backgrounds by introducing a source term, $S_{\rm inj} = \id n_{\rm inj}/\id \tau$ into the photon evolution equation. Generally speaking, the injection source term can take any number of creative forms, of which several have been extensively studied using \texttt{CosmoTherm} \citep[e.g. in][]{Chluba2015GreensII, Bolliet2020PI}. In \cite{Acharya2023SPH} and \cite{Acharya2023b}, we presented a dedicated numerical study of this effect by considering scenarios of decaying dark matter particles, and those with a power-law time-dependence. We would like to stress that some level of soft photon heating occurs even when no extra photons are introduced, as highlighted in the previous section. In addition, even in standard scenarios one can expect a noticeable amount of soft photon production, e.g., from early structure formation with radio-loud galaxies, an avenue that we plan to pursue in the future. As we show in this section, the presence of a soft photon background can have a dramatic impact on various cosmological observables. 

To illustrate this effect, we begin by studying scenarios in which a soft photon background is introduced in a quasi-instantaneous manner. This is meant more to build intuition in understanding how the various heating rates of the medium are affected, rather than to present a realistic, physically motivated scenario. To achieve this, we present results based on four representative injection redshifts, $z_{\rm inj} = 3000, \, 1000, \, 500, \, 100$, with each of these highlighting a qualitatively different epoch in the thermal history. The $z_{\rm inj} = 3000$ and $z_{\rm inj} = 1000$ scenarios capture the heating dynamics when injection takes place before and during recombination. The dark ages injection at $z_{\rm inj} = 500$ provides insight into a time when $X_{\rm e}$ is low, while at $z_{\rm inj} = 100$ we can study what happens when soft photons are sourced after $T_{\rm M}$ and $T_{\rm CMB}$ fully decouple.

We generically assume the soft photons follow a power law with spectral index given by $\gamma$. To better understand the importance of this spectral slope we compare and contrast between three choices of $\gamma$ at each injection redshift. Our choices are $\gamma = 3.0$, mimicking a free-free emission type spectrum, $\gamma = 3.6$ which is a typical synchrotron-type injection, and $\gamma = 3.3$ as an ad-hoc choice to understand intermediate slope power laws. We define the amplitude of the soft photon background relative to the CMB as $\Delta \rho/\rho$, and fix this quantity to be $10^{-6}$ for most of the examples we showcase. At late times ($z\lesssim 10^3$), it is in principle possible to take $\Delta \rho/\rho \simeq 6\times 10^{-5}$ (or even somewhat larger) without violating constraints from CMB spectral distortions, though these large amplitude injections over a quasi-instantaneous timescale lead to other complications that we discuss in the following subsections. We have found that $\Delta \rho/\rho = 10^{-6}$ provides a good benchmark to highlight the salient features of soft photon heating.

In the following subsections, we present a general prescription for the injection of soft photon backgrounds, and present our main results for quasi-instantaneous injections. After discussing these results, we highlight some subtleties related to the rather unphysical nature of quasi-instantaneous injections. We conclude by describing changes to the formalism when the soft photon background is injected solely in the Rayleigh-Jeans tail ($x \ll 1$), a scenario with specific applications to the observed radio synchrotron background.

\subsection{Generic soft photon backgrounds}
\label{sec:generic_BG}
The distinction between ``soft" and ``hard" photons is that soft photons should not be capable of exciting or ionizing neutral hydrogen in the ground state. Therefore, the injections that we perform are limited to frequencies where $x < x_{\alpha} = E_{\alpha}/k_{\rm b} T_{\rm CMB}(z)$ to ensure no hard heating takes place. Recalling that the intensity spectrum of an injected background is related to its occupation number by Eq.~\eqref{eq:Intensity-occupation} and assuming that the injection spectrum does not depend explicitly on time, we can parameterize the soft photon occupation number by
%
\begin{align}
    n_{\rm SPB}(\nu,\tau) = \zeta(\tau) \, f(\nu).
\end{align}
%
For quasi-instantaneous injections, we choose $\zeta(\tau) = A \, \Theta(\tau-\tau_{\rm inj})$, where $\Theta$ is the Heaviside step function, $\tau_{\rm inj}$ the time of injection, and $A$ is the dimensionless amplitude of the background at $\tau_{\rm inj}$ (and $\nu_{\rm cut}$, defined below). This initial amplitude can be used to normalize the total energy injection, as we will see shortly. The frequency dependence is chosen to replicate a power law background, 
%
\begin{align}
    f(\nu) = \left(\frac{\nu}{\nu_{\rm cut}}\right)^{-\gamma} \, {\rm e}^{-\nu/\nu_{\rm cut}}, \label{eq:Sinj}
\end{align}
%
where $\gamma<4$ is the spectral index (as described earlier), and $\nu_{\rm cut}$ is a high frequency cutoff which we impose to prevent the production of hard photons. A cutoff such as this could reasonably be expected in certain physical production processes (e.g. from the decay of a dark matter particle). In general, the $\zeta(\tau)$ and $f(\nu)$ terms can take any form, but we choose to focus on power law backgrounds for this work. The source term for our quasi-instantaneous injections is then given by
%
\begin{align} \label{eq:S_inj}
    S_{\rm inj}(x,\tau) = A \, \left(\frac{x}{x_{\rm cut}}\right)^{-\gamma} \, {\rm e}^{-x/x_{\rm cut}} \, \delta(\tau - \tau_{\rm inj}).
\end{align}
%
Here, we have transformed from $\nu \rightarrow x$. Additionally, $x_{\rm cut} = E_{\rm cut}/k_{\rm b}T_{\rm CMB}$ is the dimensionless cutoff scale which inherits a redshift dependence from $T_{\rm CMB}\propto (1+z)$. One can then define the rate of comoving energy injection from this source term as
%
\begin{align} 
\label{eq:src_inj_rate}
    \frac{1}{a^4} \frac{\id (a^4 \rho_{\rm inj})}{\id \tau} &= \frac{\rho_{\rm CMB}}{\pi^4/15} \int_0^{\infty} \id x \, x^3 \, S_{\rm inj}(x, \tau), \nonumber
    \\
    &= \frac{\rho_{\rm CMB}}{\pi^4/15} \, A \,x_{\rm cut}^{4}\,  \Gamma(4-\gamma) \, 
    \delta(\tau - \tau_{\rm inj})
\end{align}
%
Note that this is the total energy injection rate of photons into the background, which cannot be immediately identified with the heating rate of the hydrogen gas (for this a Compton or Bremsstahlung interaction has to occur). From here we can define the total fractional energy density increase relative to the CMB, $\Delta \rho/\rho$. For injections taking place between an initial and final redshift ($\tau(z_{\rm i}) = 0$ and $\tau(z_{\rm f}) = \tau_{\rm f}$), this is given by
%
\begin{align} \label{eq:src_frac_inj}
    \frac{\Delta \rho}{\rho} = \int_0^{\tau_{\rm f}} \id \tau \, \frac{1}{a^4 \rho_{\rm CMB}} \frac{\id (a^4 \rho_{\rm inj})}{\id \tau}.
\end{align}
%
For our quasi-instantaneous source term, assuming that $\tau_{\rm inj} < \tau_{\rm f}$, we find a simple analytic form for the normalization constant
%
\begin{align} \label{eq:A_coeff}
    A &=\frac{\pi^4}{15\, \Gamma(4-\gamma) }\, \left(\frac{k_{\rm b} T_{\rm CMB}(\tau_{\rm inj})}{E_{\rm cut}}\right)^4 \, \frac{\Delta \rho}{\rho}.
\end{align}
%
For more general time-dependence, one can simply compute the normalization condition using Eq.~\eqref{eq:src_frac_inj}, where now the explicit time-dependence of $\id \zeta(\tau)/ \id \tau$ has to be specified. The form given in Eq.~\eqref{eq:A_coeff} is particularly useful when one wishes to fix the fractional energy injection $\Delta \rho/\rho$ of a background. 

We must now specify some parameter combinations in order to illustrate various interesting cases. We set the high frequency cutoff to be $E_{\rm cut} \simeq 0.235$ eV, which corresponds to $x_{\rm cut}(z=0) = 10^3$, leaving us free from hard photons. For the injection redshifts and spectral indices, we consider $z_{\rm inj} = 3000, \, 1000, \, 500, \, 100$ and $\gamma = 3.0, \, 3.3,$ and $3.6$ in what follows. As discussed above, we fix $\Delta \rho/\rho = 10^{-6}$ and set the amplitude $A$ according to Eq.~\eqref{eq:A_coeff} for different spectral indices and injection redshifts. This allows us to showcase the importance of soft photon heating without violating constraints from CMB spectral distortions or anisotropy measurements. This also has the side effect of greatly reducing the amplitude of the soft photon background for injections occurring at lower $z_{\rm inj}$, a point which we return to at the end of this subsection.

\begin{figure*}
\centering 
\includegraphics[width=\columnwidth]{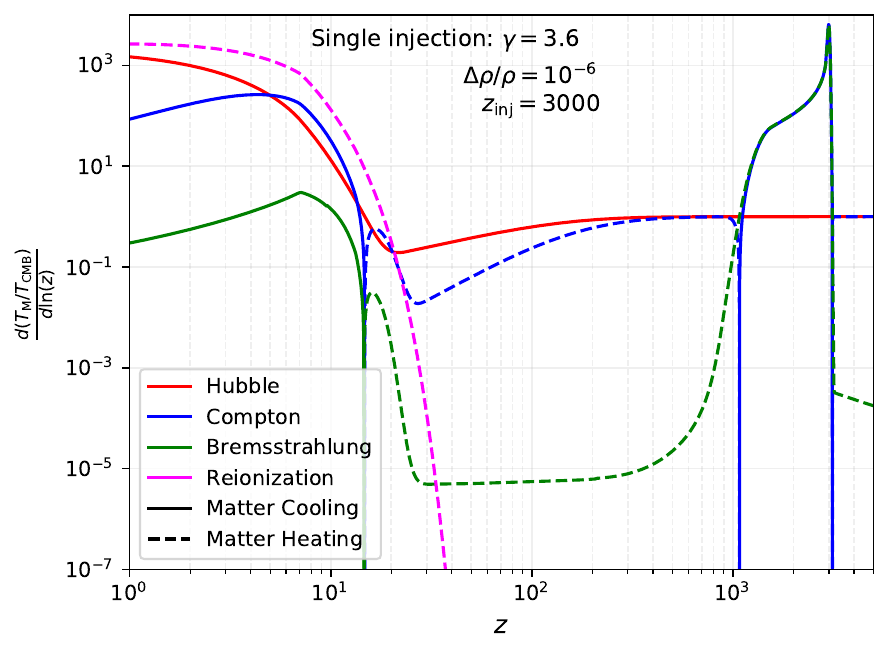}
\hspace{4mm}
\includegraphics[width=\columnwidth]{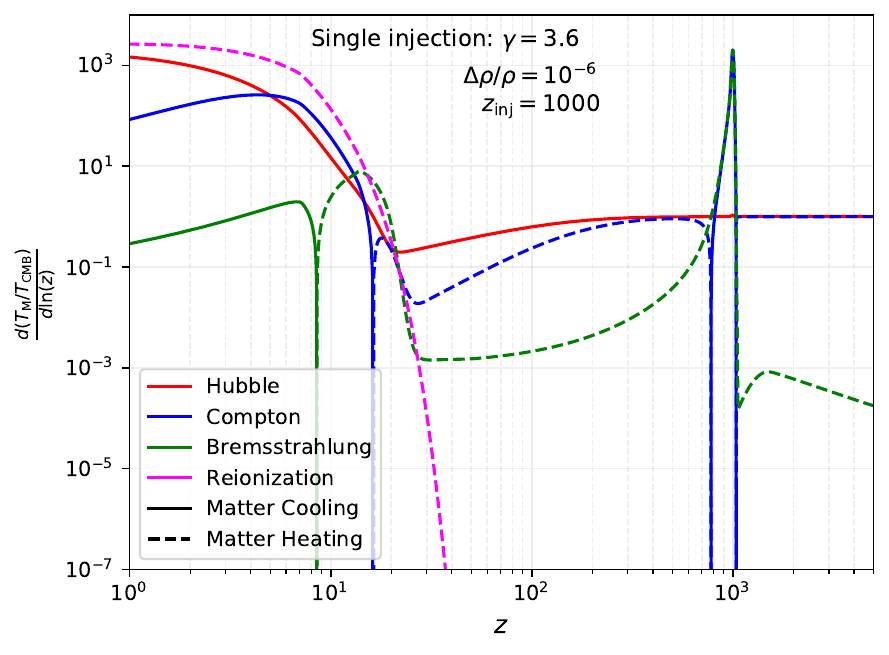}
\\[5mm]
\includegraphics[width=\columnwidth]{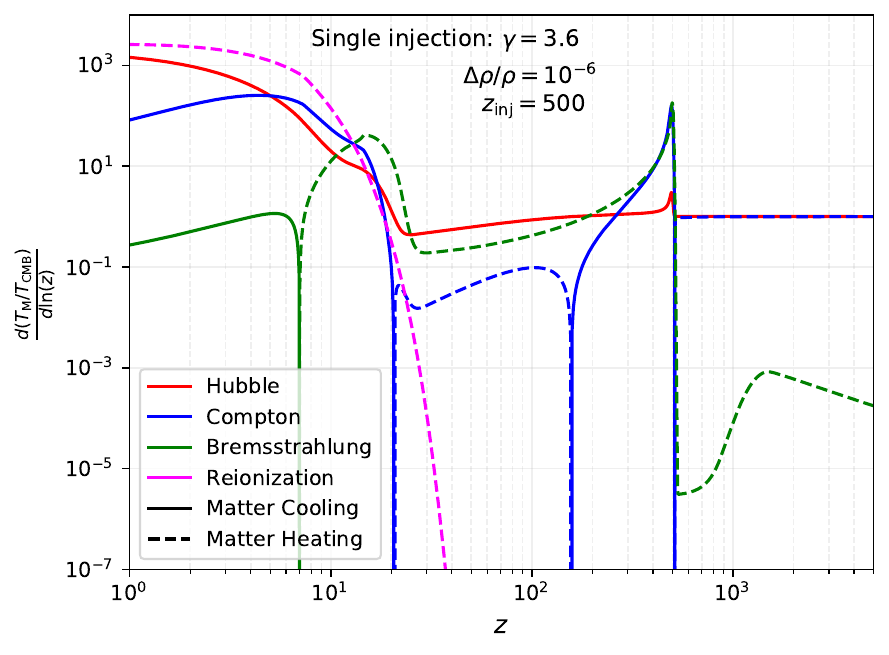}
\hspace{4mm}
\includegraphics[width=\columnwidth]{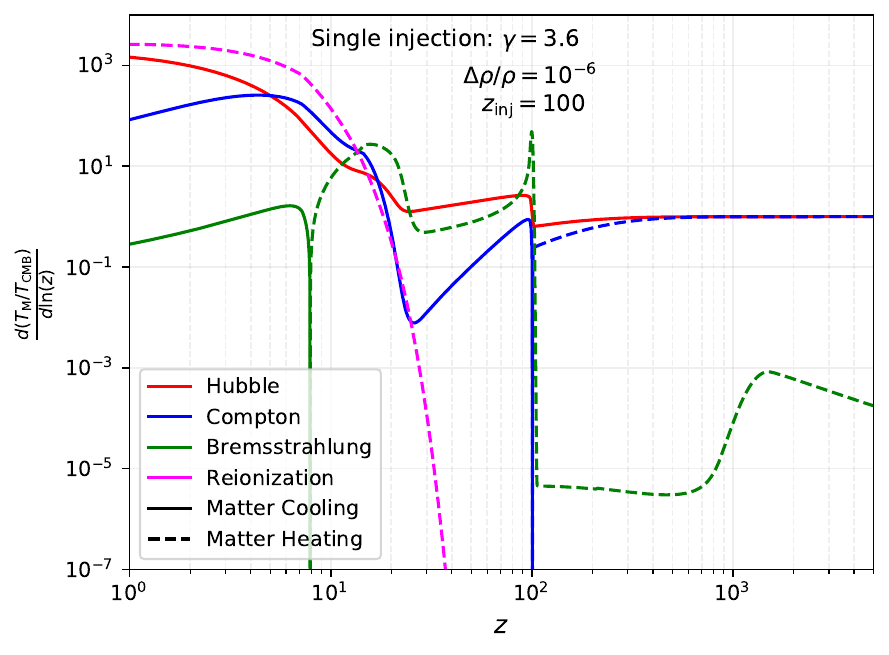}
\caption{Relative heating rates for quasi-instantaneous injections at a variety of redshifts with spectral index $\gamma = 3.6$. For pre-recombination injections, the late-time heating rates roughly match those in Fig.~\ref{fig:Vanilla-heating-rates}. Post-recombination injections lead to greatly enhanced Bremsstrahlung (soft photon) heating. When free-free heating dominates over the reionization contour at cosmic dawn ($z \simeq 20$), large corrections to the induced $\Delta T_{\rm b}$ need to be taken into account.}
\label{fig:Heating-rates-Comp}
\end{figure*}

In Fig.~\ref{fig:Heating-rates-Comp}, we present a detailed look at the heating rates from single injections of a synchrotron ($\gamma = 3.6$) spectrum at each of our target redshifts. Starting from the top left, as expected the heating rates look very similar to the vanilla $\Lambda$CDM scenario (Fig.~\ref{fig:Vanilla-heating-rates}), with the exception of the spike around $z_{\rm inj} = 3000$. At this high redshift, the pre-recombination plasma is highly efficient at absorbing low frequency photons (through free-free), while Compton scatterings quickly redistribute this newly absorbed energy back into the photon bath, leading to no net deviation between the electron and photon temperatures (though this causes a $y$-distortion with $y\simeq \frac{1}{4} \Delta \rho/\rho$). 

For injections taking place at $z_{\rm inj} \leq 1000$, the situation begins to change quite dramatically. At $z_{\rm inj} = 1000, \, 500$ after an initial boost in $T_{\rm M}$ near the injection point through soft photon heating, the relatively efficient Compton interactions quickly bring the matter temperature back to $\simeq T_{\rm CMB}$. However, the residual free-free heating rate reaches a steady-state that is significantly enhanced compared to pre-recombination injections and even exceeds the Compton heating for the case with $z_{\rm inj} = 500$. For $z_{\rm inj} = 100$, Compton interactions have almost completely frozen out and soft photon heating leads to a permanent boost in the matter temperature when compared to the case of no injections.

\begin{figure*}
\centering 
\includegraphics[width=\columnwidth]{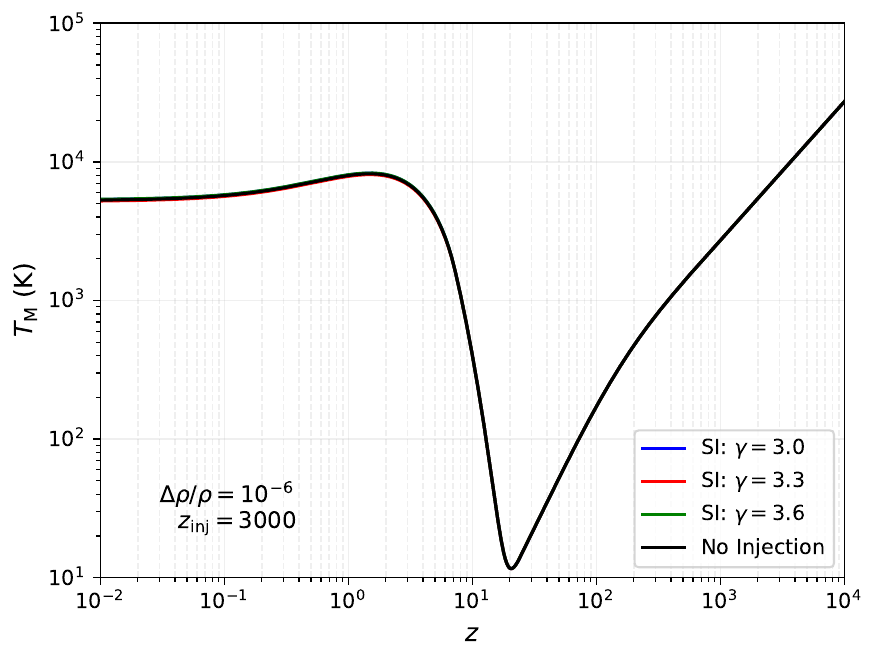}
\hspace{4mm}
\includegraphics[width=\columnwidth]{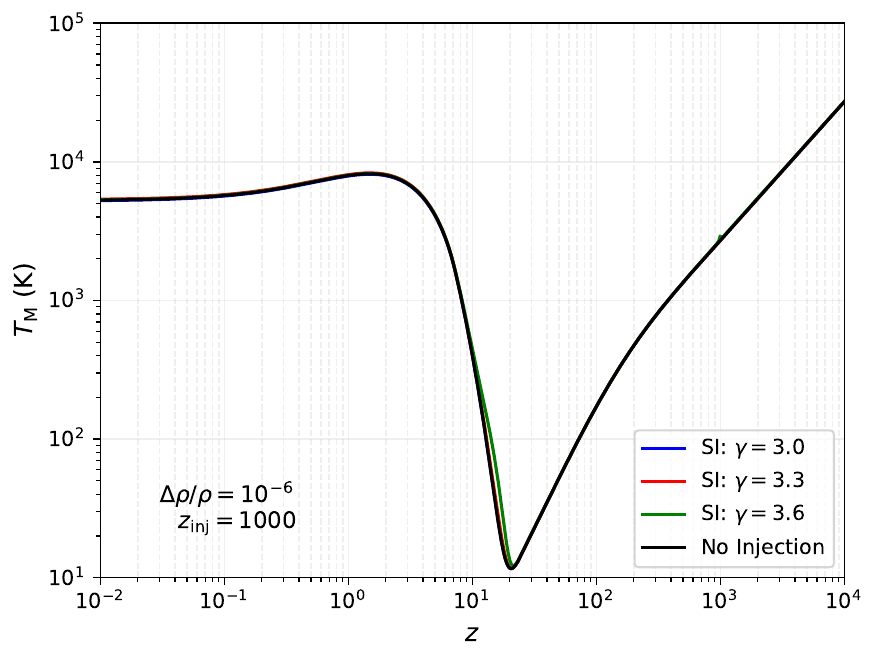}
\\[5mm]
\includegraphics[width=\columnwidth]{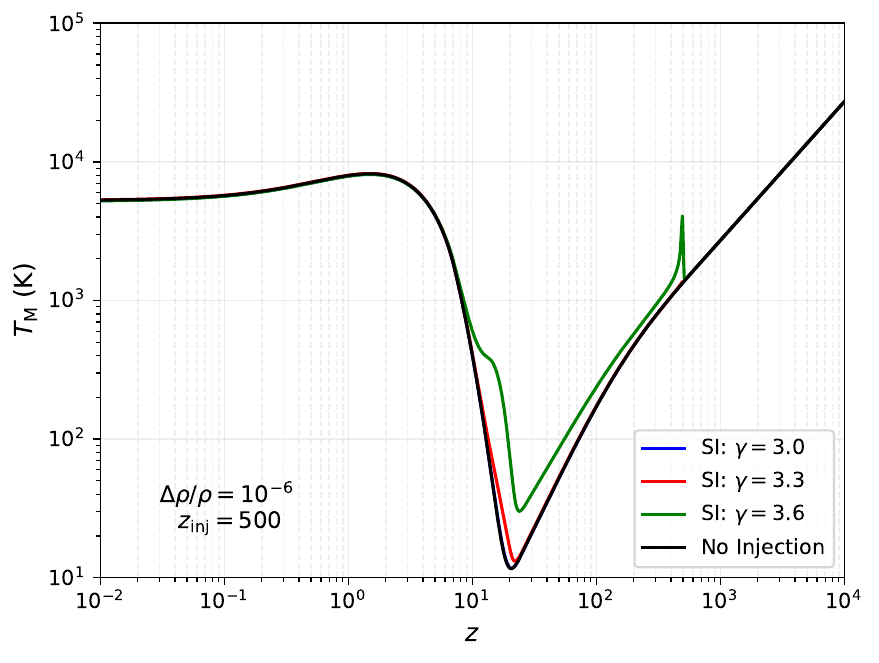}
\hspace{4mm}
\includegraphics[width=\columnwidth]{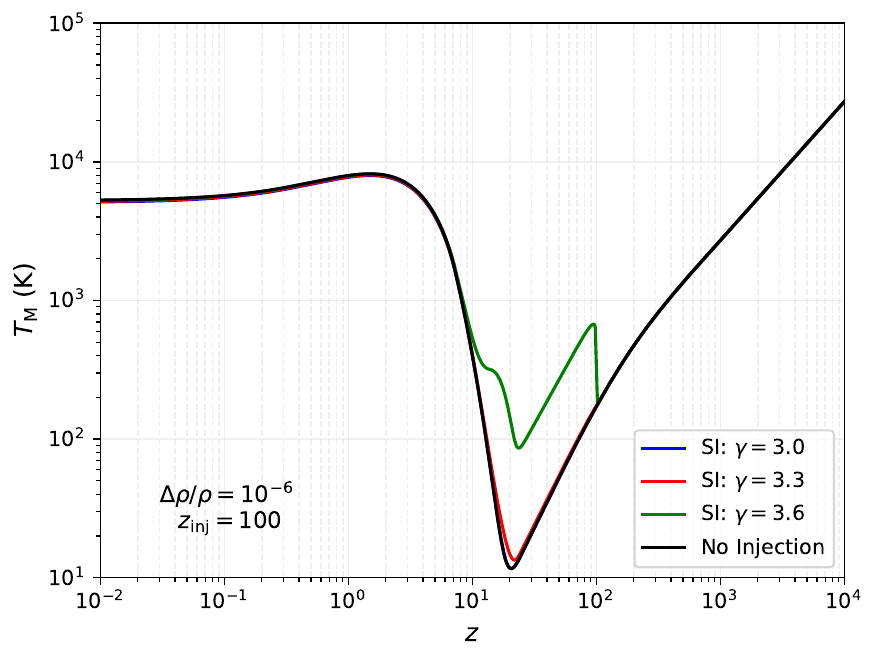}
\caption{Evolution of the matter temperature for free-free, synchrotron, and $\gamma = 3.3$ spectral indices. The inefficiency of Compton scattering at $z \lesssim 150$ leads to a large offset in the matter temperature for synchrotron-type injections. Deviations from the no injection scenario at $z \simeq 20$ indicate the presence of soft photon heating relevant to global $21$cm observations. }
\label{fig:Long-range-TM-Comp}
\end{figure*}
\begin{figure*}
\centering 
\includegraphics[width=\columnwidth]{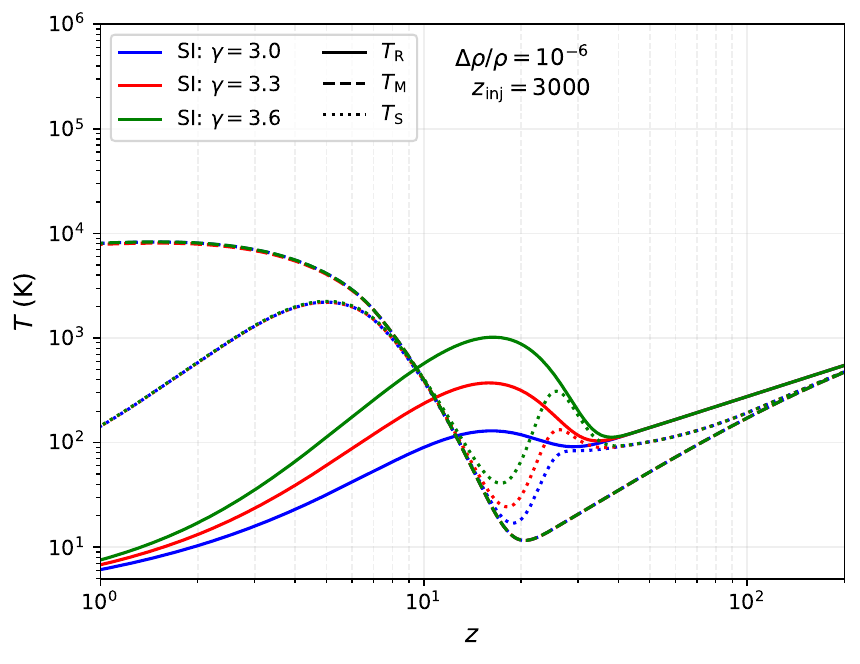}
\hspace{4mm}
\includegraphics[width=\columnwidth]{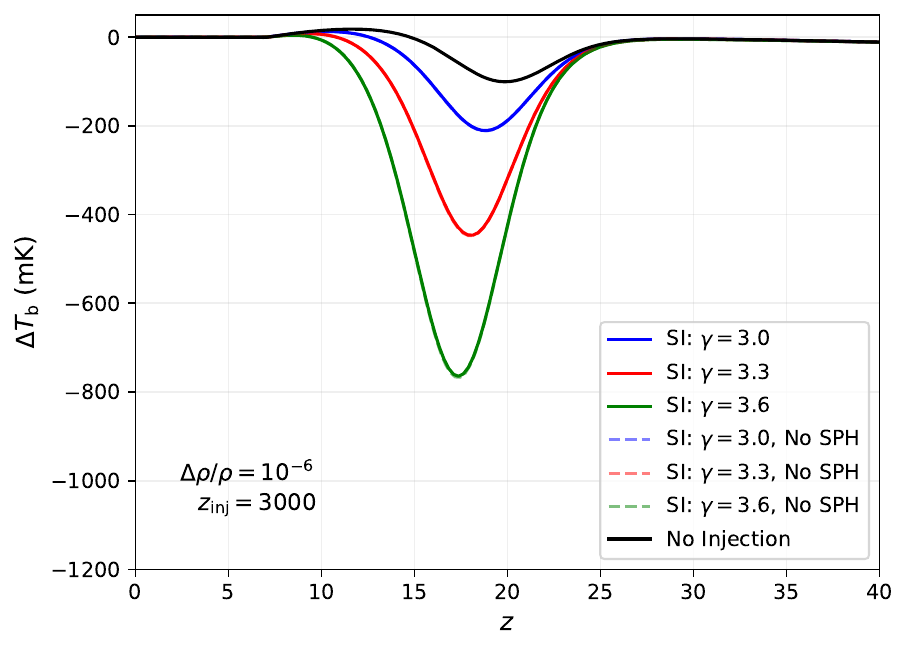}
\\[5mm]
\includegraphics[width=\columnwidth]{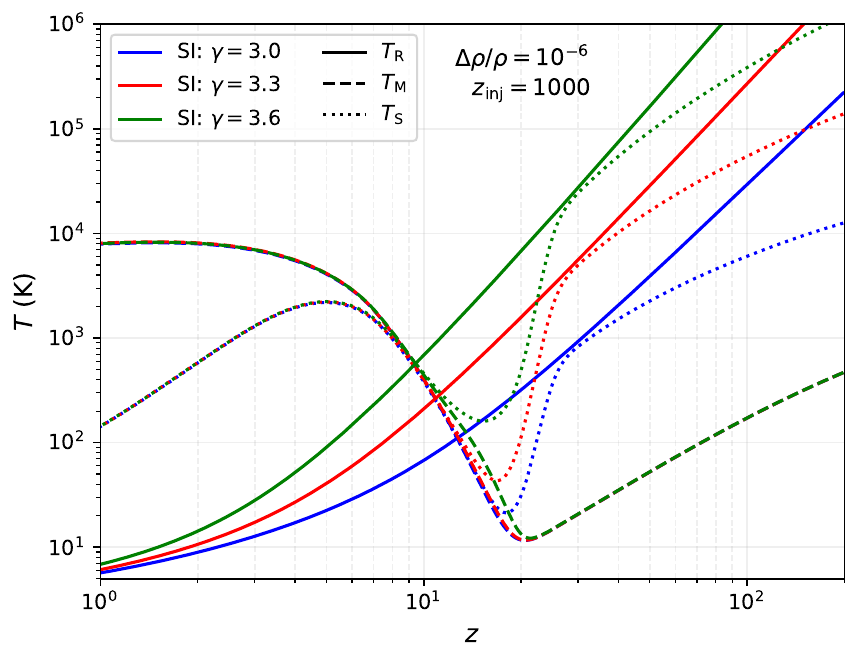}
\hspace{4mm}
\includegraphics[width=\columnwidth]{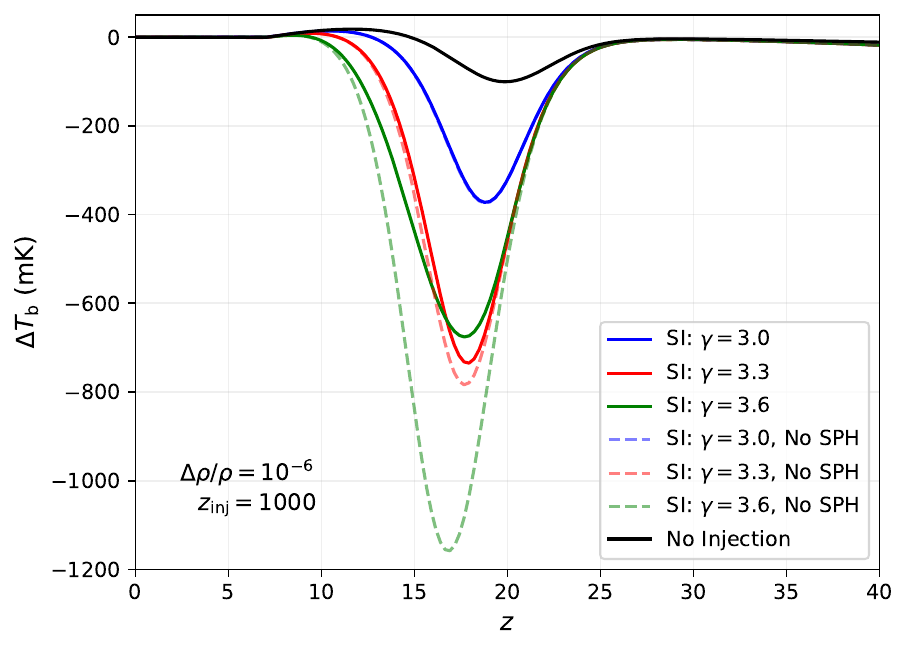}
\\[5mm]
\includegraphics[width=\columnwidth]{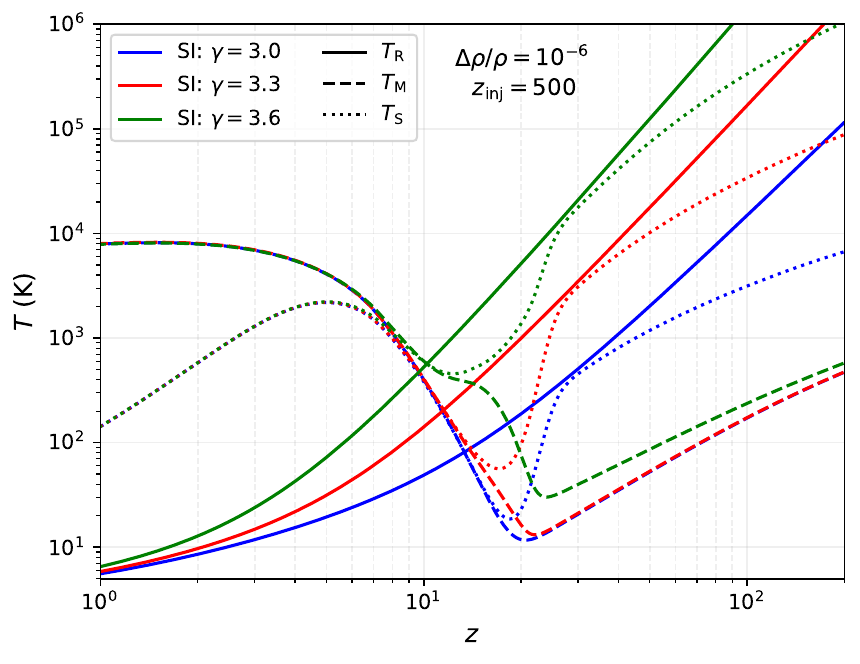}
\hspace{4mm}
\includegraphics[width=\columnwidth]{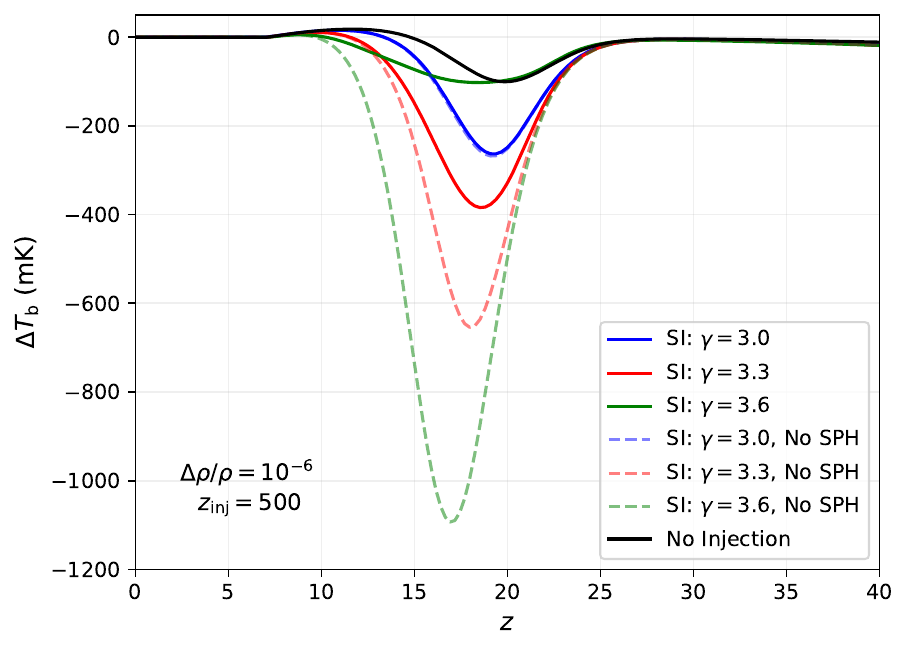}
\caption{Left panels: A breakdown of the radiation, matter, and spin temperatures for different injection redshifts spectral indices. Right panels: The differential brightness temperature in the presence of each type of soft photon background, both with soft photon heating, and by neglecting it by setting $\id T_{\rm ff}/\id z = 0$ in Eq.~\eqref{eq:T_M}. When a soft photon spectrum permeates the background, soft photon heating can only be neglected if the background is sourced in the pre-recombination epoch, or if the spectral index is $\gamma \leq 3.0$. In all other circumstances, heating through free-free absorption will generically source large corrections to the differential brightness temperature at cosmic dawn.}
\label{fig:dTb-21cm-Comp-a}
\end{figure*}
\begin{figure*}
\centering 
\includegraphics[width=\columnwidth]{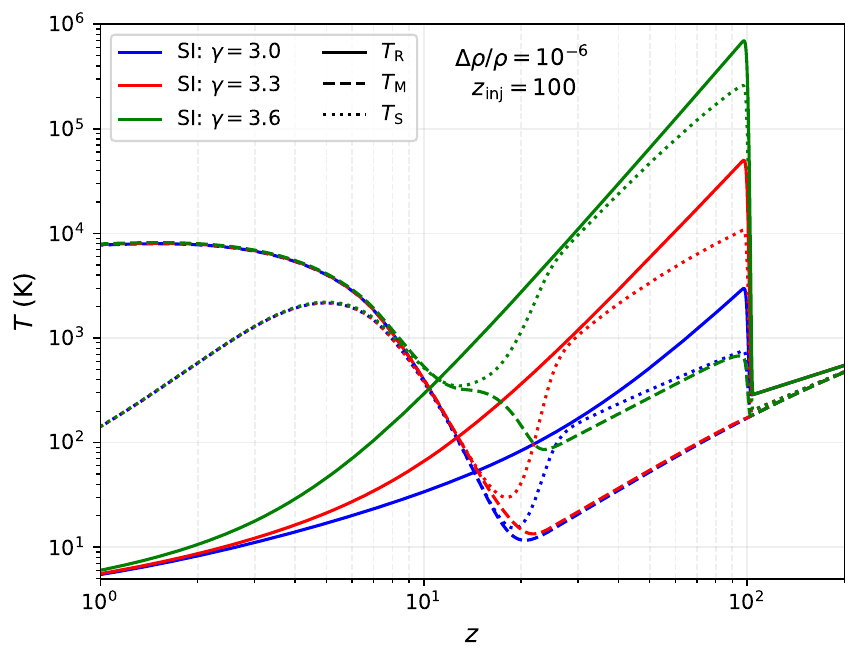}
\hspace{4mm}
\includegraphics[width=\columnwidth]{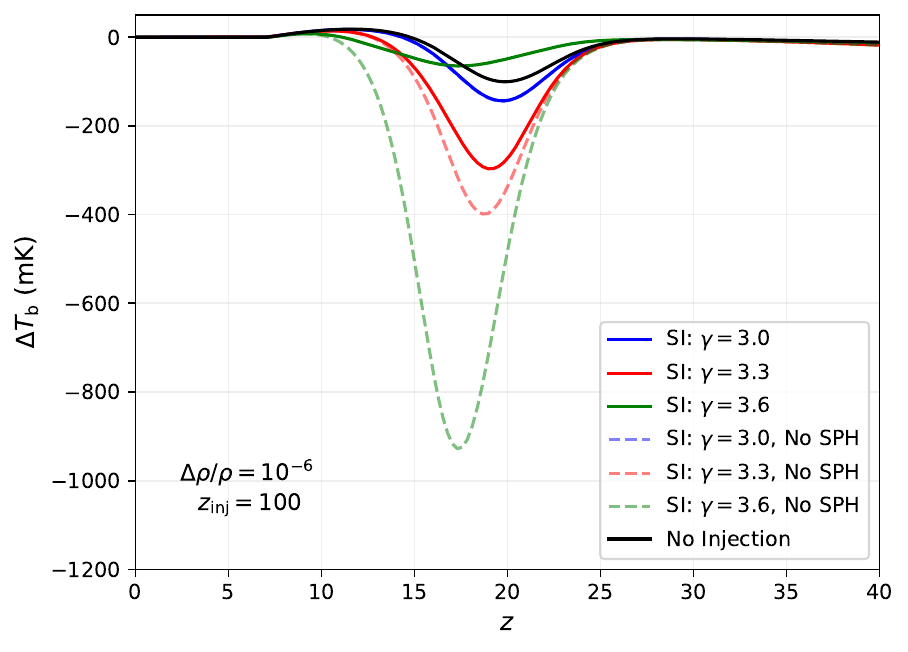}
\caption{The same as Figure~\ref{fig:dTb-21cm-Comp-a}, but for $z_{\rm inj} = 100$. The unphysical spike comes from the non-adiabatic injection and immediate reprocessing of the soft photon background.}
\label{fig:dTb-21cm-Comp-b}
\end{figure*}

Most important, however, is the fact that the secondary Bremsstrahlung bump around redshift of $z \simeq 20$ can dominate over the heating sourced by reionization. For pre-recombination injections, the free-free optical depth is extremely high for low-frequency photons. This implies that the tail of an injected soft photon background is quickly truncated at a relatively high frequencies. For injections after recombination, this initial truncation takes place at much lower frequencies. When reionization begins to turn on, $X_{\rm e}$ rapidly increases and a second period of soft photon heating occurs as free-free absorption once again becomes efficient. Therefore, a generic consequence of post-recombination soft photon injections is this two stage heating process, once near $z_{\rm inj}$, and once at the onset of reionization, around cosmic dawn.

The plots in Fig.~\ref{fig:Long-range-TM-Comp} show the deviation in the matter temperature over a wide range of redshifts for $\gamma = 3.0, \, 3.3, \,$ and $3.6$ with different injection times. The initial spike in the $z =  1000, \, 500$ scenarios happens as a result of the quasi-instantaneous injection, after which $T_{\rm M}$ quickly relaxes back to the CMB temperature due to Compton interactions. For $z_{\rm inj} = 100$, the initial temperature jump remains frozen in as the hydrogen continues to cool adiabatically. The sharp increase in $T_{\rm M}$ around $z \simeq 20$ comes from our reionization module, which can also trigger a second phase of soft photon heating for the post-recombination injections as mentioned above.

We present a detailed breakdown of the evolution of radiation, matter, and spin temperatures for our different case studies in the left hand panels of Figs.~\ref{fig:dTb-21cm-Comp-a} and \ref{fig:dTb-21cm-Comp-b}, focusing on the range $1 \leq z \leq 200$. The amplitude and time dependence of the differential brightness temperature is governed by the contrast between the solid ($T_{\rm R}$) and dotted ($T_{\rm S}$) lines. We present our computed $\Delta T_{\rm b}$ curves in the right hand panels, comparing both with the vanilla $\Lambda$CDM setup with no soft photon injections, as well as the predicted curves expected when one neglects soft photon heating in the presence of such a background. These ``No SPH" curves are computed by simply setting $\id T_{\rm ff}/\id z = 0$ in Eq.~\eqref{eq:T_M}.

It is immediately evident from the first plot that for $z_{\rm inj} = 3000$, soft photon heating has a negligible impact on the late time evolution of the matter temperature, as it does not deviate from the no-injection case presented in Fig.~\ref{fig:dTb-Vanilla}. The radiation temperature at $21$cm frequencies begins to receive a boost around $z=30$ due to the fact that initial truncation at $z_{\rm inj}$ has redshifted down to the relevant wavelengths. Injections at $z \gg 3000$ will not see any such bump, as the entirety of the injected background at those redshifts would be quickly absorbed. The lack of free-free heating is also evident in the top-right panel of Fig.~\ref{fig:dTb-21cm-Comp-a}, where the differential brightness temperature with and without a $\id T_{\rm ff}/\id z$ term completely overlap. This result matches nicely with the conclusions of e.g. \citet{Feng2018}, who showed that large enhancements to $\Delta T_{\rm b}$ are possible in the presence of a radio background. 

The remainder of the plots in Figs.~\ref{fig:dTb-21cm-Comp-a} and \ref{fig:dTb-21cm-Comp-b} tell a qualitatively different story to what previous literature \citep{Feng2018, Ewall-Wice2018, Fialkov2019} originally concluded. Deviations of the matter temperature away from $\Lambda$CDM are observed due to the second phase of soft photon heating for $z_{\rm inj} \leq 1000$. Extra thermal energy in the gas due to soft photon heating acts to pump the occupation of the spin triplet state, giving an additional boost to $T_{\rm S}$ at cosmic dawn. This increase in spin temperature leads to a net reduction in the amplitude of $\Delta T_{\rm b}\propto x_{\rm H}(1-T_{\rm R}/T_{\rm S})$, which in most circumstances cannot reasonably be neglected. The centre of the $\Delta T_{\rm b}$ absorption trough can also experience a slight offset depending on the exact redshift that the second phase of soft photon heating starts to dominate.

Plots in the right hand column of Figs.~\ref{fig:dTb-21cm-Comp-a} and \ref{fig:dTb-21cm-Comp-b} show the magnitude of these corrections for single injections with different spectral indices and injection redshifts. It is clear that the effect is most prominent for backgrounds with a steeper spectral index, produced after recombination. This is intuitive, as it is due to the fact that a higher relative fraction of the soft photon background is absorbed through free-free heating when the spectrum is steeper. 

For a free-free power-law ($\gamma = 3.0$) injection, soft photon heating is rather negligible, while for a synchrotron-like ($\gamma = 3.6$) spectra, the effect can reduce the amplitude of $\Delta T_{\rm b}$ even below the predicted $\Lambda$CDM value, making it much harder to detect. The synchrotron case is particularly interesting, due to indications from ARCADE-2 \citep{Fixsen2011excess} and LWA \citep{DT2018} that an unexplained radio synchrotron background may be present. If such a background does exist and was produced at $z \gtrsim 20$, our calculations indicate that the amplitude of the global $21$cm differential brightness temperature could be far less than the standard prediction ($|\Delta T_{\rm b}| \ll 100$ mK).

Finally, we should address why the amplitude of energy injections seems to be decreasing with lower $z_{\rm inj}$. This can be seen most easily by looking at the peak heating rates at $z_{\rm inj}$ in Fig.~\ref{fig:Heating-rates-Comp}, or the steadily decreasing amplitudes of $\Delta T_{\rm b}$ or $T_{\rm R}$ for $z_{\rm inj} = 1000, \, 500,$ and $100$ in Figs.~\ref{fig:dTb-21cm-Comp-a} and \ref{fig:dTb-21cm-Comp-b}. This stems from the fact that we have chosen to fix the ratio $\Delta \rho/\rho$, which implies that the initial amplitude of soft photon backgrounds injected at later times is suppressed, as can be seen in Eq.~\eqref{eq:A_coeff}. If one instead chooses to fix $\Delta \rho$, this subtlety can be avoided, but comes at the cost of introducing a more complicated mapping to e.g. CMB spectral distortion constraints which are sensitive to the fractional energy injection relative to the CMB.

\subsection{Energetic photon backgrounds and collisional ionizations}
\begin{figure*}
\centering 
\includegraphics[width=\columnwidth]{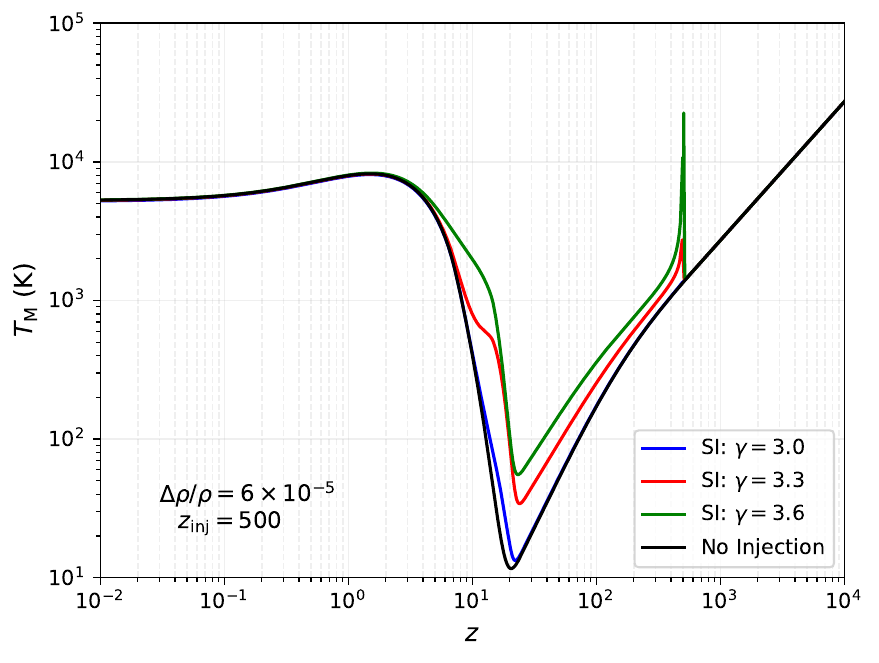}
\hspace{4mm}
\includegraphics[width=\columnwidth]{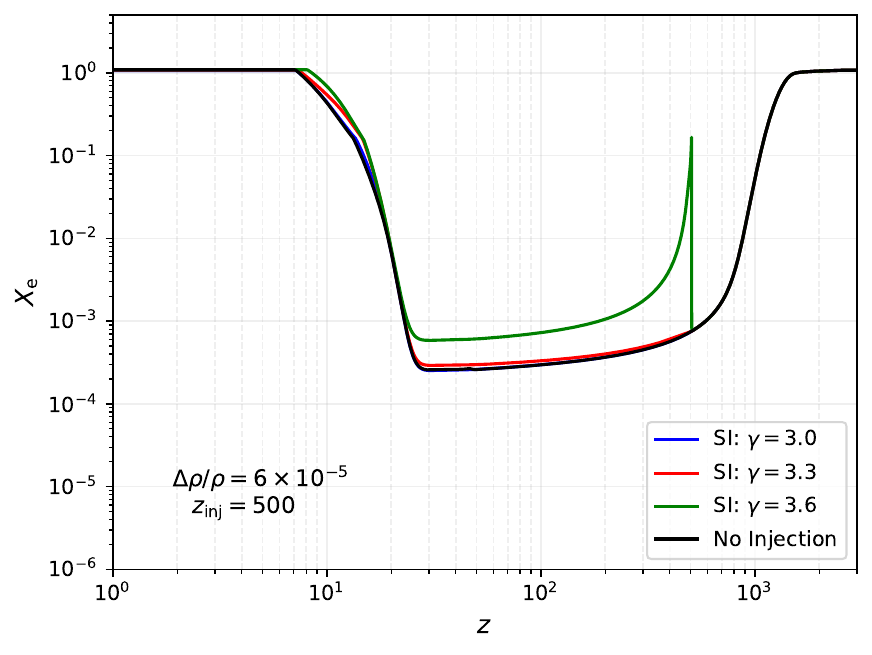}
\\[5mm]
\includegraphics[width=\columnwidth]{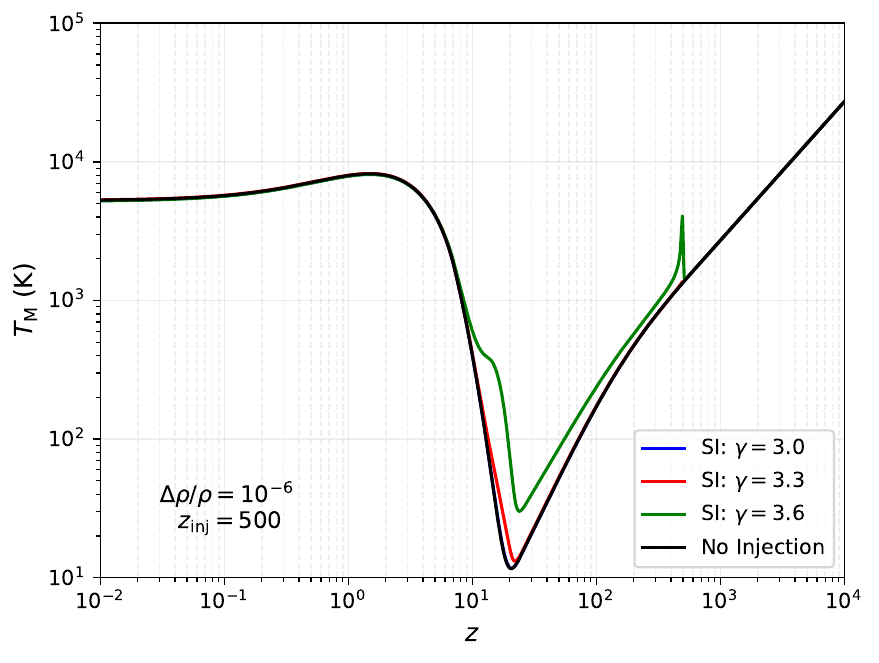}
\hspace{4mm}
\includegraphics[width=\columnwidth]{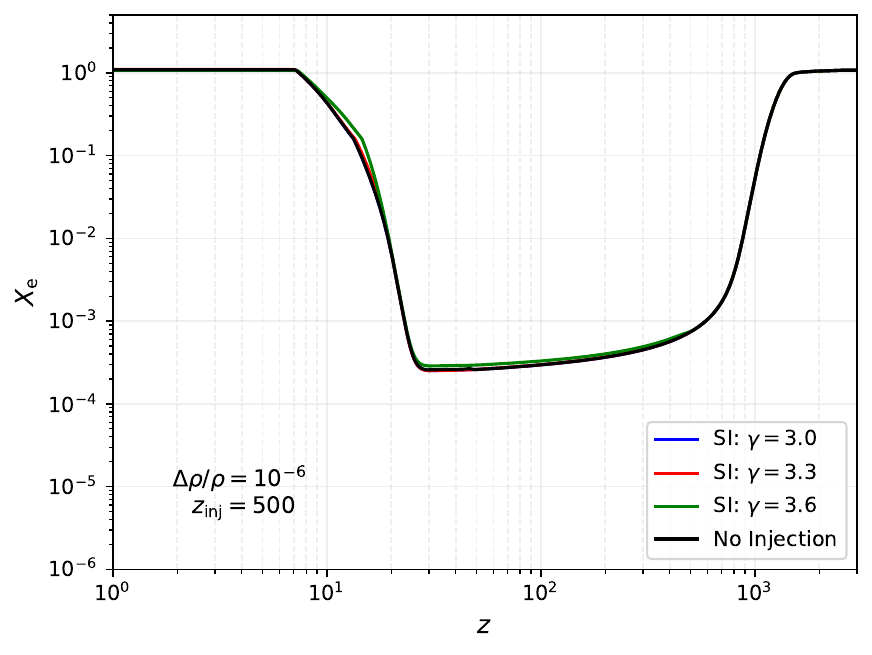}
\caption{Top panels: Evolution of the matter temperature and ionization fraction over a wide range of redshifts for a single injection at $z_{\rm inj}$ with $\Delta \rho/\rho = 6\times 10^{-5}$. Bottom panels: Same as the top but for a less energetic spectrum with $\Delta \rho/\rho = 10^{-6}$. For the more energetic case, the matter temperature briefly spikes to $T_{\rm M} \gtrsim 10^4$ K, inducing efficient collisional ionizations that strongly perturb the $X_{\rm e}$ history.}
\label{fig:Collisional-Ionization}
\end{figure*}

Before moving on we would like to highlight an additional complication that can arise when performing quasi-instantaneous injections with a brighter background spectrum (i.e., a larger amplitude $A$). In the upper two panels of Fig.~\ref{fig:Collisional-Ionization}, we show the matter temperature and ionization fraction evolution with redshift for a single injection at $z_{\rm inj} = 500$, and fractional energy density $\Delta \rho/\rho = 6\times 10^{-5}$. This value of $\Delta \rho/\rho$ was chosen as it saturates the bound on exotic energy injections set by the \COBEF measurements on CMB spectral distortions.

From the the upper left panel, we can see that for $\gamma = 3.6$, the initial spike in the matter temperature peaks at $T_{\rm M}(z_{\rm inj}) \simeq 3\times 10^4$ K. As mentioned earlier, the threshold for when collisional ionizations become important is roughly $T_{\rm M} \gtrsim 10^4$ \citep[e.g.,][]{Chluba2015PMF}, which we clearly exceed for a brief time in this scenario. The result of crossing this threshold is a simultaneous spike in the $X_{\rm e}$ fraction, which soon relaxes to a new steady state value a sizeable fraction higher than the standard $\Lambda$CDM value. For reference we plot the $\Delta \rho/\rho = 10^{-6}$ results in the bottom two panels, showing that such a large perturbation in $X_{\rm e}$ is avoided for lower energy injections.

While it is obviously important to not disturb the ionization fraction of the universe in such a dramatic fashion, we would like to stress that this situation is an artifact of the quasi-instantaneous nature of these soft photon injections used in the illustrations here. For more physical scenarios, such as the continuous injections we consider in Section \ref{sec:cont_inj}, the soft photon background builds up more slowly over time. This avoids large instantaneous spikes in $T_{\rm M}$ by smoothing out the initial phase of soft photon heating over a large range of redshifts. Therefore, it is reasonable to take $\Delta \rho/\rho = 6\times 10^{-5}$ for continuous injection scenarios without violating constraints on $X_{\rm e}$. We note that in the post-recombination era, this bound can still be exceeded as not all the injected energy is reprocessed into $y$-type distortions. General scenarios therefore required a detailed parameter study \citep{Acharya2023b}.

\subsection{A radio synchrotron background source term}
In the previous subsections, we studied the impacts of soft photon heating when one fixes the fractional energy release ($\Delta \rho/\rho)$ from an injected background. This was useful to understand the relevance of different spectral indices and injection redshifts. Here, we briefly illustrate how to study these effects for a specified form of a radio synchrotron background, i.e., when the amplitude $A$ of the background is fixed \textit{a priori}. This is motivated by measurements from the ARCADE-2 and LWA experiments which indicate that an unexplained radio synchrotron background may exist. 

Radio backgrounds such as this appear as deviations from the CMB spectrum in the Rayleigh-Jeans tail ($x < 1$). They are typically parameterized by a radio brightness temperature, $T_{\rm amp}(\tau)$, pivot frequency, $\nu_0$, and a (temperature) spectral index, $\beta$, in the form
%
\begin{align}
    T_{\rm RSB}(\nu,\tau) = T_{\rm amp}(\tau) \left(\frac{\nu}{\nu_0}\right)^{-\beta}.
\end{align}
%
We note that $\gamma = \beta + 1$ relates the temperature spectral index with the $\gamma$ used in the previous subsections.The occupation number of the radio background can be related to this parameterization through
%
\begin{align}
    n_{\rm RSB}(x,\tau) = \frac{1}{x} \, \frac{T_{\rm RSB}(x,\tau)}{T_{\rm CMB}(\tau)}.
\end{align}
%
If we once again assume that the radio synchrotron background is injected quasi-instantaneously at some $\tau_{\rm inj}$, we find a source term similar to Eq.~\eqref{eq:Sinj}, namely 
%
\begin{align} \label{eq:S_RSB}
    S_{\rm RSB} = 
    \tilde{A}\,\left(\frac{x}{x_{\rm cut}}\right)^{-\gamma} \, {\rm e}^{-x/x_{\rm cut}} \, \delta(\tau - \tau_{\rm inj}).
\end{align}
%
Here, $\tilde{A} = (k_{\rm b}T_{\rm amp}/h\nu_0)\,(x_{\rm cut}/x_0)^{-\gamma}$, which allows us to make use of the results from Sect.~\ref{sec:generic_BG} with $A\rightarrow \tilde{A}$. Again, we modulate the spectrum by imposing an exponential cutoff in frequency for $x_{\rm cut} < 1$ to ensure we remain in the Rayleigh-Jeans tail. 
\begin{figure*}
\centering 
\includegraphics[width=\columnwidth]{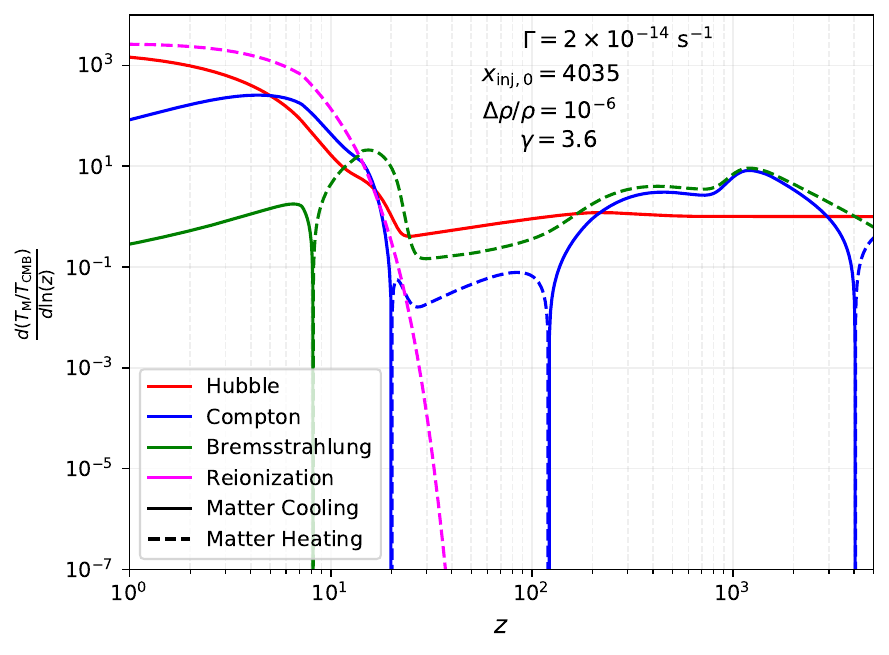}
\hspace{4mm}
\includegraphics[width=\columnwidth]{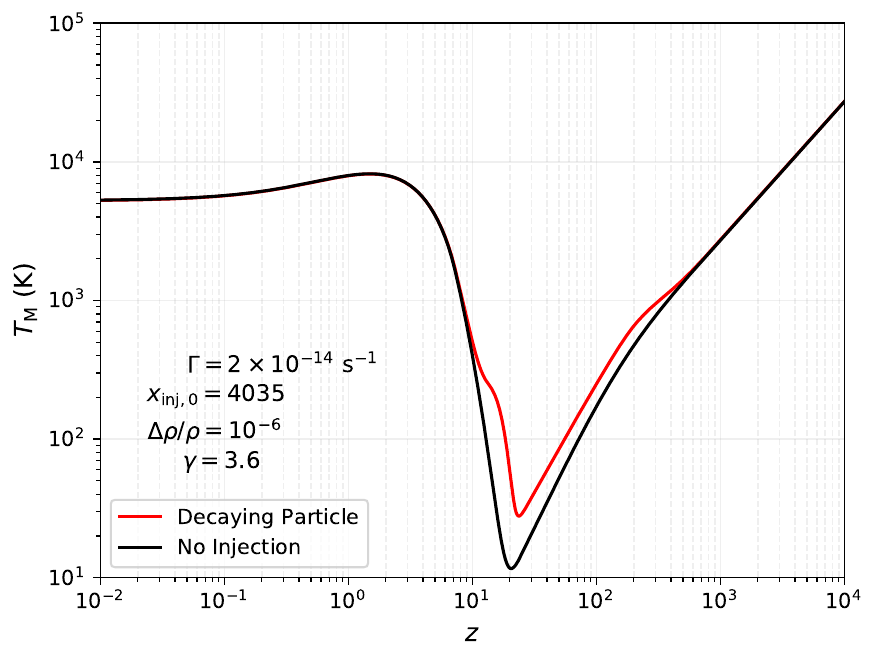}
\\[5mm]
\includegraphics[width=\columnwidth]{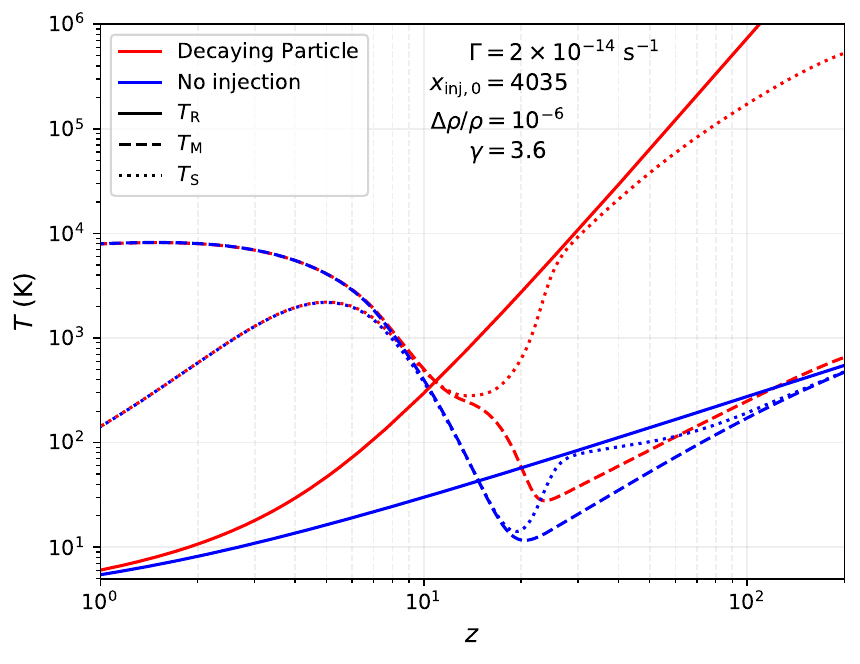}
\hspace{4mm}
\includegraphics[width=\columnwidth]{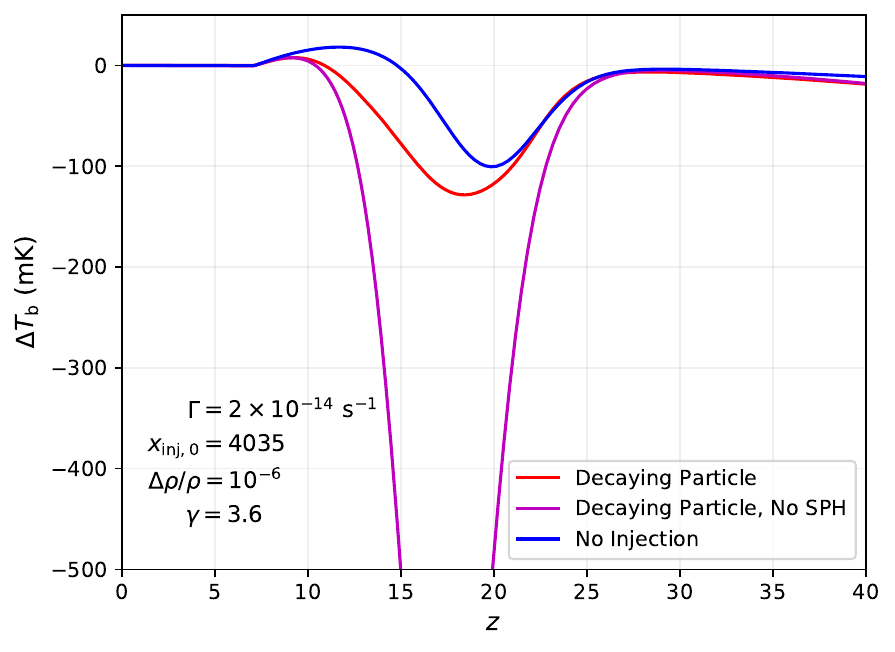}
\caption{An overview of the heating mechanisms at play in the decaying dark matter scenario. The specific parameters that we choose correspond to a model which provides a good fit to the observed radio synchrotron background, as discussed in \citet{Acharya2023SPH}. A decay rate of $\Gamma = 2\times 10^{-14} \, {\rm s}^{-1}$ corresponds roughly to a redshift of $z \simeq 420$, after which soft photon production no longer occurs. Qualitatively, the results are similar to the case of quasi-instantaneous injection at $z_{\rm inj} = 500$, which we considered above.}
\label{fig:DP-overview}
\end{figure*}
%
\begin{figure*}
\centering 
\includegraphics[width=\columnwidth]{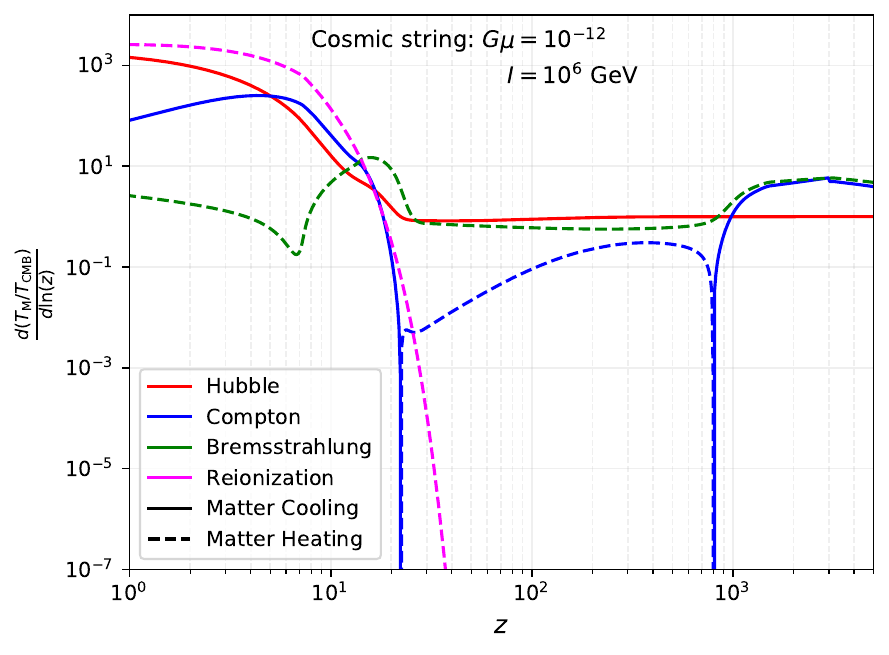}
\hspace{4mm}
\includegraphics[width=\columnwidth]{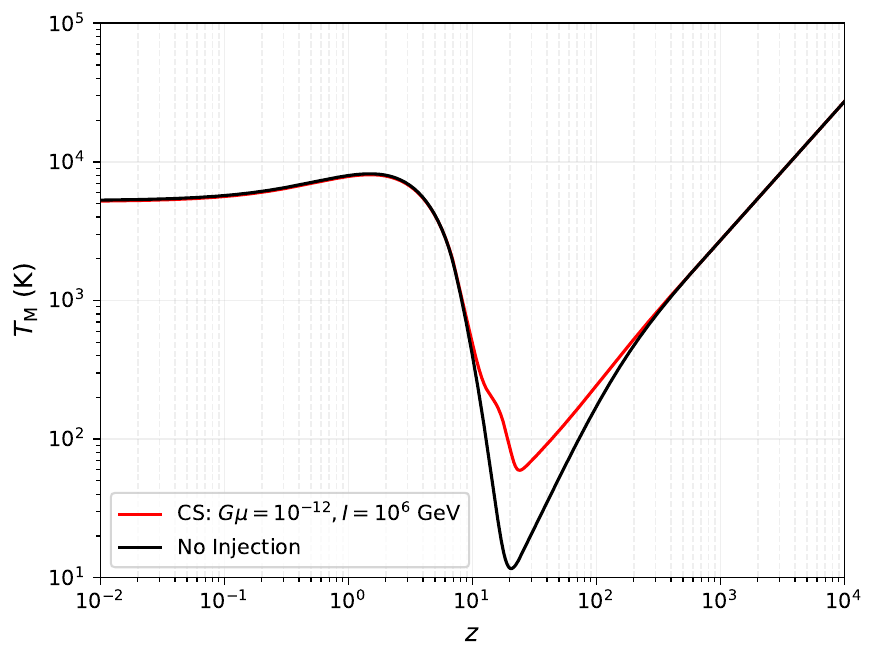}
\\[5mm]
\includegraphics[width=\columnwidth]{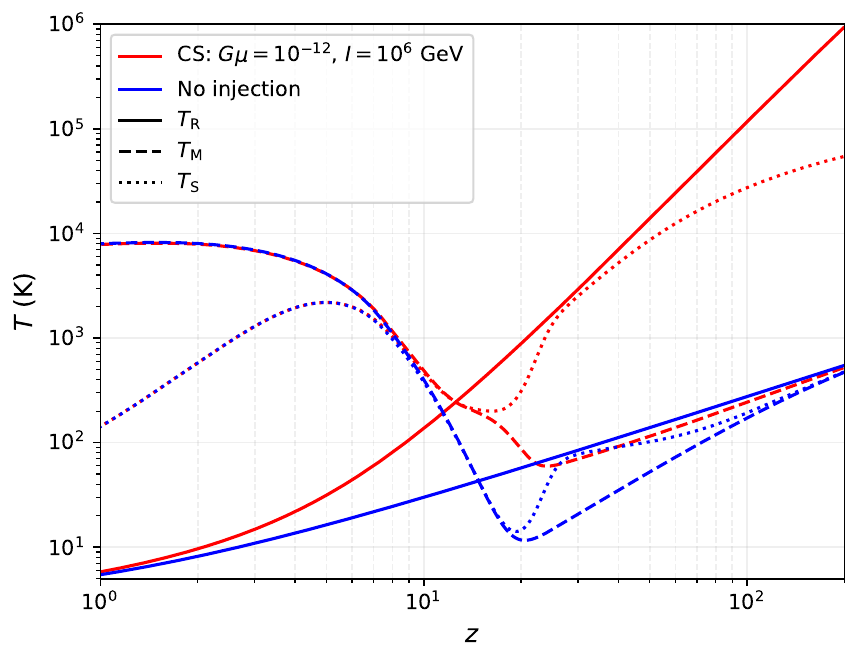}
\hspace{4mm}
\includegraphics[width=\columnwidth]{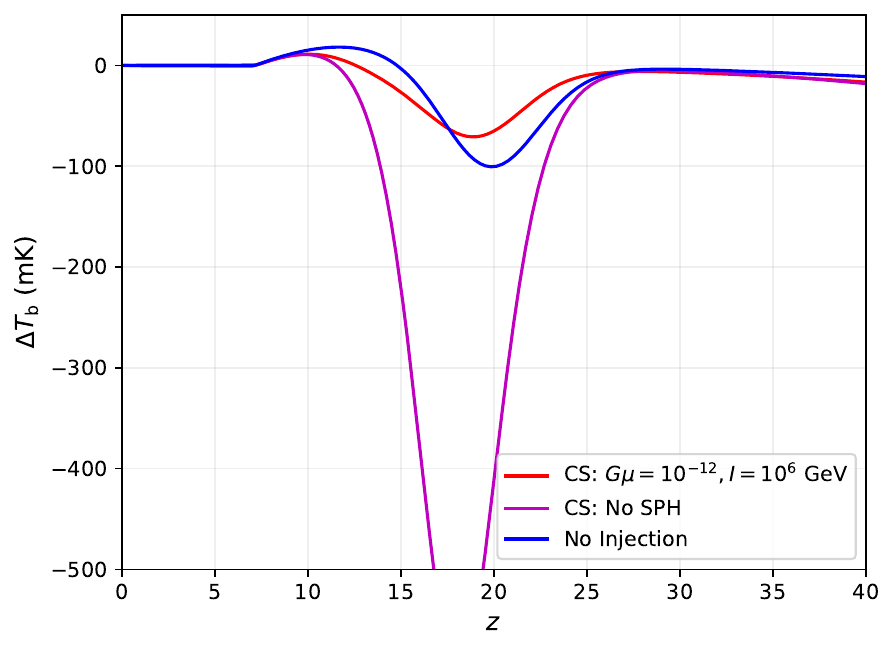}
\caption{The same breakdown as in Fig.~\ref{fig:DP-overview}, but for a superconducting cosmic string scenario with $G\mu = 10^{-12}$ and $\mathcal{I} = 10^6$ GeV. In contrast to the other cases considered in this work, a superconducting cosmic string network is capable of producing an energetic spectrum of soft photons at all redshifts, leading to consistent free-free heating. The parameters chosen in this depiction represent a model that is also quite a convincing fit to the ARCADE-2 and LWA data regarding the radio synchrotron background \citep{Cyr2023CSS}.}
\label{fig:CS-overview}
\end{figure*}
%

This injection model consists of five parameters $\tilde{A}$, $\gamma$, $x_0$, $x_{\rm cut} (< 1)$ and $\tau_{\rm inj}$. As a simple example, let us return to the radio synchrotron background that has been reported by the ARCADE-2 and LWA experiments. In Eq. (6) of \citet{Acharya2023SPH}, we explored a power-law fit to this background, which can be translated into our parameterization yielding $\gamma = 3.555$, $x_0 (z = 0) \simeq 0.0176$, and $\tilde{A} \simeq 25.6$ in the case where $x_0 = x_{\rm cut}$.

Strictly speaking, the injection source term given in Eq.~\eqref{eq:S_RSB} is valid in the limit $x_{\rm cut} < 1$, and useful when a radio brightness temperature, $T_{\rm RSB}(\nu,\tau)$ is given. In contrast, the source given in Eq.~\eqref{eq:S_inj} is valid for $E_{\rm cut} \lesssim 10$ eV. The accompanying Jupyter Notebook incorporates both of the formalisms for these source terms and allows one to quickly and easily understand which types of low frequency backgrounds will yield appreciable soft photon heating.

\section{Continuous Injections}
\label{sec:cont_inj}
While quasi-instantaneous injections of a soft photon background provided us with a useful tool to understand the underlying dynamics of free-free heating, they are perhaps not the most physically relevant scenarios. To supplement our understanding, we will briefly consider two scenarios in which a low-frequency photon spectrum is built up over time. 

\subsection{Decaying dark matter}
The first scenario we wish to consider is that of a decaying dark matter particle (DP). In this setup, we imagine that there exists some fraction $f_{\rm dm}$ of the dark matter which is unstable and decays into a broad spectrum of photons\footnote{The role of dark matter decays into two photons has been extensively studied in \citet{Bolliet2020PI}.}. The energy injection rate is given by
%
\begin{align} \label{eq:DP-inj-rate}
    \frac{\id \rho_{\rm DP}}{\id t} = f_{\rm dm} \, \rho_{\rm dm} \, \Gamma \, {\rm e}^{-\Gamma t},
\end{align}
%
where $\rho_{\rm dm}$ is the dark matter energy density, and $\Gamma$ is the decay rate (or inverse lifetime) of the unstable dark matter particle. Similar to the previous section, the spectrum of decay products is assumed to follow a power law with a high frequency cutoff,
%
\begin{align}
    I_{\rm DP} = I_{\rm 0, DP} \, x^{3-\gamma} {\rm e}^{-x/x_{\rm inj}}
\end{align}
%
with $x_{\rm inj}=x_{\rm inj, 0}/ (1+z)$. Studying this scenario allows us to directly extend the results from the previous section to a setup with a non-trivial time dependence. To fix the total energy release $\Delta \rho/\rho$ (and thereby determine $f_{\rm dm}$), we have to compute the normalization of the injection spectrum (by integrating over $x^3 \id x$) and also the total energy release integral to determine the constant $f_{\rm dm}$. This process can be carried out numerically. Further details of this model can be found in \citet{Acharya2023SPH}, where we performed an in-depth numerical analysis using \texttt{CosmoTherm}. 

To illustrate the effects of soft photon heating, we select the parameters $\gamma = 3.6$, 
$\Delta \rho/\rho = 10^{-6}$, $x_{\rm inj,0} = 4035$, and $\Gamma = 2\times 10^{-14} \, {\rm s}^{-1}$ which corresponds to a decay redshift of $z_{\rm dec} \simeq 420$, after which very few soft photons are produced. This particular combination of parameters is chosen such that no hard photons are emitted, and because the part of the spectrum that survives to late times offers a good fit to the ARCADE-2 radio synchrotron background data. 

A breakdown of the decaying particle scenario is presented in Fig.~\ref{fig:DP-overview}. The top left panel shows that at early times, the soft photon injection follows a power law in redshift as indicated by the steadily increasing Bremsstrahlung heating. This heating rate is slightly reduced at recombination when the free electron fraction drops, but continues heating at a roughly constant rate until $z \simeq z_{\rm dec}$ when the majority of the dark matter particles have decayed away. At this point, most of the soft photon injection stops, and the situation becomes qualitatively very similar to the single injection scenario with $z_{\rm inj} = 500$ as presented in Fig.~\ref{fig:Heating-rates-Comp}.

The remainder of the plots in Fig.~\ref{fig:DP-overview} also tell a similar story to what we saw with quasi-instantaneous injections near $z = 500$, showing a strong departure in both the spin and $\Delta T_{\rm b}$ temperatures from the $\Lambda$CDM case. This leads us to the somewhat obvious conclusion that the most potent effects of soft photon heating on $21$cm observations come from late time free-free absorptions. Therefore, if a soft photon source turns off at some point (as is the case in both the quasi-instantaneous and decaying particle scenarios), the relevant observational effects come from the two heating bumps at $z_{\rm inj}$ or $z_{\rm dec}$, as well as near cosmic dawn. For the decaying particle scenario discussed here we also explicitly compared our simplified treatment with the full numerical result computed using {\tt CosmoTherm}, finding very good agreement. 

\subsection{Superconducting cosmic strings}
A setup in which soft photons are produced at all redshifts is given by the superconducting cosmic string scenario, originally studied in \citet{Cyr2023CSS, Cyr2023a}. In this picture, cosmic strings are formed at the interface of a cosmological phase transition at an arbitrarily high redshift, and endowed through subsequent symmetry breakings with superconductive properties (see \citet{Witten1984,Ostriker1986} for seminal examples). On relatively short timescales after formation of the strings, a scaling network is achieved, with $\mathcal{O}(1)$ long strings running through the Hubble patch at any given redshift \citep{Kibble1980, Vilenkin2000}. 

Alongside these long strings, a distribution of smaller, sub-Hubble loops is constantly sourced as the network evolves \citep{Vanchurin2005, Martins2005, Ringeval2005, Blanco-Pillado2013}. These smaller loops oscillate and are thought to slowly decay away into both gravitational waves \citep{Vachaspati1984} and electromagnetic radiation \citep{Vilenkin1986, Vachaspati2008, Cai2012}. The fraction of energy going into gravitational waves is controlled by the so-called string tension parameter $G\mu$, where $G$ is Newton's constant and $\mu \sim \eta^2$ where $\eta$ is the energy scale of the phase transition which produced the cosmic strings. Similarly, the strength of electromagnetic emission is controlled by the amplitude of the current, $\mathcal{I}$, running along the string.

The string-loop produced gravitational waves provide a possible explanation to the recently detected stochastic background of detected by the pulsar timing array consortium \citep{NANOGrav2023, EPTA2023, Ellis2023}. Interestingly, the spectrum of electromagnetic radiation released contains a large fraction of soft photons as can be seen in Fig. 12 of \citet{Cyr2023a}. Importantly, since a string network such as this would be expected to survive until the present day, the persistent decay of cosmic string loops ensures that new soft photons are constantly being pumped into the background. Therefore, the soft photon heating process is extended gradually over a wide range of redshifts. Importantly, the second bump in the free-free heating rate at cosmic dawn is still present (this is simply due to an increase in $X_{\rm e}$ around this time).

The details of the exact injection rate and fractional $\Delta \rho/\rho$ are somewhat complicated, so we refer the interested reader to our dedicated papers on the subject. In \citet{Cyr2023CSS} we found that a string tension of $G\mu \simeq 10^{-12}$ and current of $\mathcal{I}\simeq 10^6$ GeV\footnote{Forgive our diversion to natural units for this brief explanation.} provided an impressive fit to the radio synchrotron background data, similar to the decaying dark matter scenario discussed above. We study the same model here in Fig.~\ref{fig:CS-overview}.

As expected in the upper left plot of this figure, the free-free heating rate is remarkably constant at early times, suffering only a small step down in amplitude over the course of recombination. This rather democratic heating can be appreciated by comparing the $T_{\rm M}$ plot in the upper right with that of the decaying particle scenario. In the decaying particle setup, $T_{\rm M}$ is boosted around $z = 400$, before soft photon heating turns off and the temperature decays adiabatically with the usual $(1+z)^2$ scaling. In contrast, the consistent heating from cosmic strings never allows this Hubble cooling to become dominant, and the matter temperature at all times until cosmic dawn follows a more photonic $(1+z)$ scaling. This is also clear by comparing the slope of the dashed red curve in the bottom right plot with that of the ``No injection" radiation temperature.

As expected, soft photon heating plays a major role in dampening the amplitude of $\Delta T_{\rm b}$ for this model. In fact, prior to our first papers discussing soft photon heating, this model of superconducting strings was ruled out as it produced a $\Delta T_{\rm b}$ far in excess of the EDGES lower bound of $-500$ mK \citep{Brandenberger2019}. The fact that soft photon heating has relaxed the parameter space of these models so significantly is quite exciting in light of the fact that they are capable of providing a very convincing fit to the observed radio synchrotron background \citep{Cyr2023CSS}.

\section{The 21cm power spectrum}
\label{sec:level6}
In this section, we demonstrate the effects of soft photon heating on the 21cm fluctuation power spectrum. Substituting the expression for $T_{\rm S}$ in expression of $\Delta T_{\rm b}$ as in Sec. \ref{sec:level2}, we find
\begin{equation}
    \begin{aligned}
    \Delta T_{\rm b}&=27x_{\rm H}(1+\delta)\left(\frac{1+z}{10}\frac{0.15}{\Omega_m h^2}\right)^{1/2}\left(\frac{\Omega_b h^2}{0.023}\right)
    \\ &\qquad \times \left(\frac{H}{{\rm dv_r/dr}+H}\right)\left(\frac{x_c+x_{\alpha}}{x_{\rm R}+x_c+x_{\alpha}}\right)
    \left(1-\frac{T_{\rm R}}{T_{\rm K}}\right)\hspace{0.2cm} {\rm mK},
\end{aligned}
\label{eq:T_b1}
\end{equation}
In the standard scenario, $T_{\rm R}=T_{\rm CMB}$ and $x_c \ll x_{\alpha}$ at $z\lesssim 30$. Ignoring redshift space distortions, fluctuations in the brightness temperature are sourced by fluctuations in the density contrast ($\delta$), as well as due to the spatially varying Lyman-alpha intensity, and from fluctuations in $T_{\rm K}$ (from spatially varying adiabatic and X-ray heating). Furthermore, during the reionization epoch, there are also fluctuations from the inhomogeneous and evolving ionization fraction.  

To compute the 21cm power spectrum, one needs to rely on simulations \citep{Iliev2006,Mellema2006} which can be computationally expensive. Semi-numerical codes such as \cite{MFC2011,21cmfast2020,Fialkov2015,Munoz21cm2023} have been developed to speed up the computation so that one can efficiently scan over a realistic range of astrophysical parameters. In this work, we have used {\tt 21cmFAST} to highlight the effect of soft photon heating for the case of a continuous injection from dark matter decay, with the same parameter values as considered in Sec. \ref{sec:cont_inj}.

For this work, we use a 100 Mpc simulation box with 100 grid points which may not be not realistic but suffices to showcase the importance of soft photon heating. We turn on the {\it USE-TS-FLUCTUATION} flag which then performs spin temperature fluctuations. The spin temperature, given in Eq.~\eqref{eq:Tspin}, is a function of both the radiation and kinetic temperatures, and is computed at each grid point in the simulation box.

We compute the source term $T_{\rm R}$ due to the non-thermal radio background such that $T_{\rm R}=T_{\rm CMB}+\Delta T$. We then plug the extra source term into the definition of $T_{\rm R}$ in {\tt 21cmFAST}. Similarly, we add this source term to the $T_{\rm K}$ (equivalently $T_{\rm M}$) equation given above. In our computation, this source term is assumed to be position independent and thus only depends on redshift. This allows us to simply shift the temperature at each grid point in the simulation box by a redshift-dependent value. In reality, there can be spatial fluctuations in the radio background, which are expected to induce additional heating. This directly affects $T_{\rm S}$ and the local value of $T_{\rm R}$, causing additional changes to the ratio $\delta [T_{\rm R}/T_{\rm S}]\approx \frac{\bar{T}_{\rm R}}{\bar{T}_{\rm S}}\left[\frac{\delta T_{\rm R}}{\bar{T}_{\rm R}}-\frac{\delta T_{\rm S}}{\bar{T}_{\rm S}} \right]$. At leading order, these contributions are expected to compensate each other; however, we do not go into these complications here. 

We also note that these source terms are computed using our fiducial reionization model which may not precisely match the output of {\tt 21cmFAST} or other such codes. However, we expect this difference to not change the qualitative features of soft photon heating, which is the effect we are aiming to highlight here. We defer a more consistent calculation, involving interfacing our code with 21cm codes, to the future. Furthermore, we have also taken care to not include the non-thermal radio photons in the Comptonization equation [Eq. \eqref{eq:T_M}] between the CMB photons and background electrons.  

\begin{figure}
\centering 
\includegraphics[width=\columnwidth]{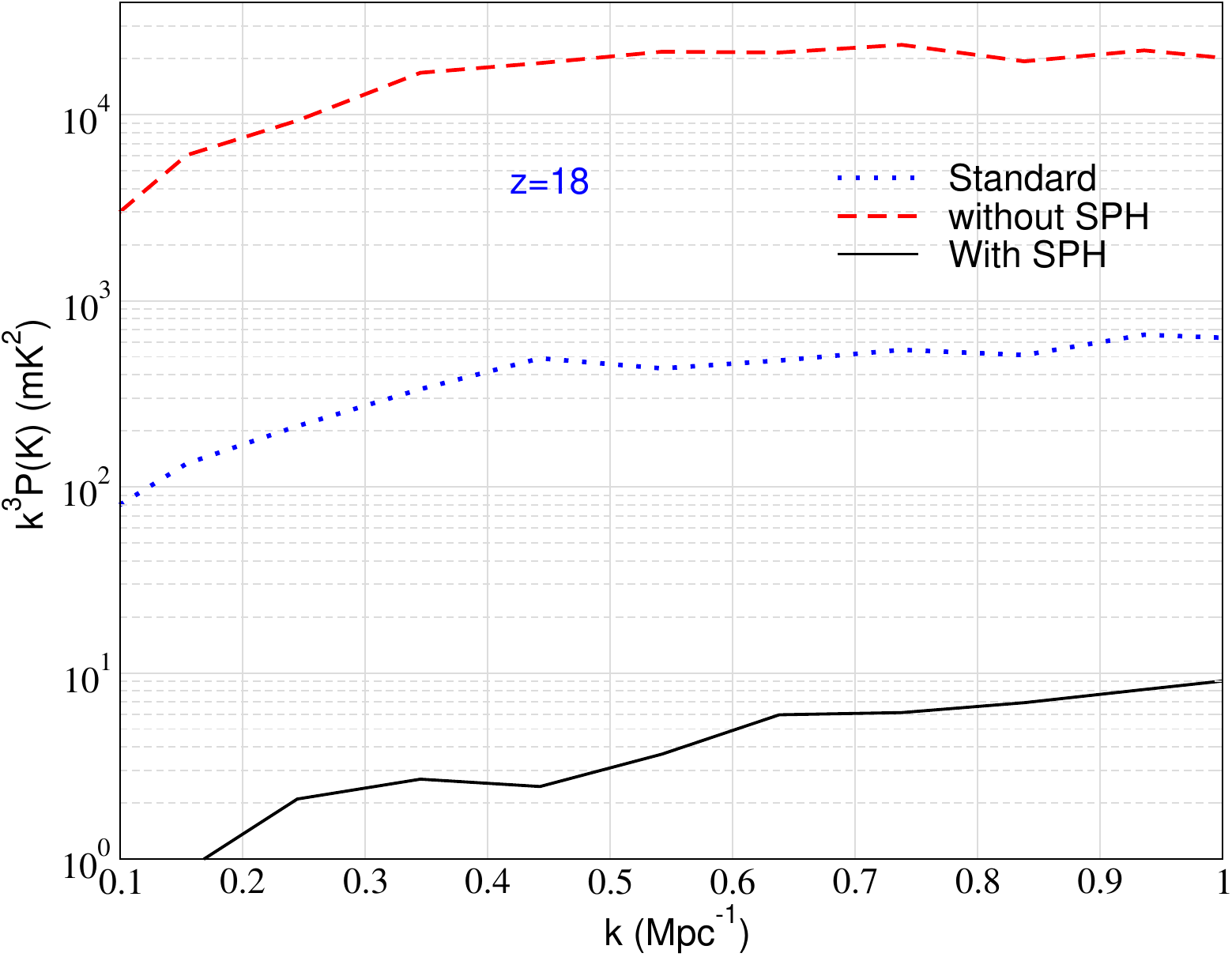}
\caption{The power spectrum for 21cm brightness fluctuations at $z=18$ for cases with and without the soft photon heating due to the presence of radio background. We consider the case with $x_{\rm inj,0}=4035$ and $\Gamma=2\times 10^{-14}$ s$^{-1}$ of our model which can produce the observed radio synchrotron background.}
\label{fig:Pk_21cmfast}
\end{figure}

In Fig. \ref{fig:Pk_21cmfast}, we show the results (at $z = 18$) for the 21cm power spectrum in a decaying dark matter scenario with a decay rate of $\Gamma = 2\times 10^{-14} \, {\rm s}^{-1}$, and a high frequency cutoff in the spectrum at $x_{\rm inj,0} = 4035$. We have used {\tt powerbox} \citep{powerbox2018} to compute the power spectrum. We see that the predicted power spectrum changes by orders of magnitude when we include the effects of soft photon heating. The 21cm fluctuations measure the temperature contrast between a hot object in a cold background or vice versa. By increasing the radio background (and neglecting soft photon heating), this contrast as well as the fluctuation signal increases. When we include soft photon heating, the overall contrast is reduced due to the increased background matter temperature. This can even push the fluctuation signal below the standard case in which there are no primordial radio backgrounds. This computation showcases the importance of including soft photon heating within the analysis of the 21cm power spectrum such as in \cite{HERA2022, HERA2022b}.  

While the discussion in this work assumes the presence of a radio background, it is expected that there are localized sources of radio photons such as radio galaxies. Recently, the authors in \cite{SBFR2024} studied the effect of radio galaxies on the 21cm power spectrum. The radio photons emitted by galaxies boost the power spectrum just as showcased in Fig. \ref{fig:Pk_21cmfast}, making them more detectable (or one can put strong upper limits on their abundance in case of non-detection). Since radio galaxies are believed to be formed by accreting black holes, it is reasonable to believe that they emit a broad spectrum of energetic photons, in addition to radio photons, which can ionize their surrounding. Following our work, one expects the radio photons to be absorbed and heat their immediate surrounding. This can suppress the 21cm signal around the location of such galaxies, and in some cases the absorption signal may switch to an emission signal depending upon the physical details. Using a simple estimate, the optical depth for radio photons turns out to be
\begin{equation}
    \tau_{\rm ff}(x,z=20)=0.32\left(\frac{10^{-6}}{x}\right)^2\left(\frac{X_{\rm e}}{0.001}\right)^2\left(\frac{g_{\rm ff}}{5}\right)\left(\frac{R}{\rm 1 Mpc}\right),
\end{equation}
where $R$ is the distance that photons travel form source of origin and we have assumed matter temperature to be 10 K at $z=20$ which follows from standard model. It is interesting to note that the optical depth is of the order of unity on Mpc scales for photons with frequency $x=10^{-6}-10^{-5}$ which contribute dominantly to soft photon heating. This would modify the powerspectrum at $k=0.1-1$ Mpc$^{-1}$ when taken into account. Implementing this effect into codes like {\tt 21cmFAST} should be similar to the treatment of X-ray photons. We plan to follow this interesting avenue in future. 

We would like to stress that in our simple demonstrations we have only considered the build-up of an excess radio background and its associated soft photon heating. Physical processes which produce higher energy quanta can undergo more traditional ``hard" photon heating, which in some cases can dominate over the soft photon heating we have discussed here. In this case,  one can expect the effect on the global 21cm signal and its fluctuations to be more more modest, since the ratio $T_{\rm R}/T_{\rm S}$ may stay more moderate. While the soft photon heating process clearly demonstrates this important interplay, from a more general perspective it is thus very important to take the correlation of heating and enhancement of a soft photon background into account. In other words, the same radio galaxies that could cause the buildup of a potential radio background will also heat the medium, with the balance of these two processes being important. A more detailed treatment of radio photon sources and heating is thus necessary and an important future direction for future work.

\section{Discussion and Conclusions}
\label{sec:level7}
In this article, we have presented a comprehensive look into the underlying physical mechanism that mediates soft photon heating. The upshot of this effect is that in the presence of a low-frequency photon background, free-free absorption cannot be ignored, as it can lead to large corrections to the gas temperature $T_{\rm M}$. In this work we have highlighted how these corrections can cause leading order effects in $21$cm cosmology by disrupting both the power spectrum and the global signal of the differential brightness temperature, $\Delta T_{\rm b}$. We would like to stress that this effect is universal, making it crucial to include as part of the standard toolbox for the computation of theoretical predictions and forecasting within the $21$cm cosmology community.

After performing a brief overview of other relevant heating mechanisms in $21$cm cosmology, we presented the coupled set of integro-differential equations that must be solved in order to properly incorporate the effects of soft photon heating, namely, Eqs.~\eqref{eq:dn-evolution}, \eqref{eq:T_M}, and \eqref{Eq:recombination}. The implementation of this effect is currently fully treated in the numerical thermalization code, \texttt{CosmoTherm} \citep{Chluba2011therm}, and work is in progress that will bring this functionality to other codes such as \texttt{DarkHistory} \citep{Liu2019, LQRS2023I, Liu2023II, Sun2023} in the near future. In the meantime, we have released a simple standalone Jupyter Notebook that can be used to explore the impact of soft photon heating for soft photon backgrounds using the injection source terms given in Eqs.~\eqref{eq:S_inj} and~\eqref{eq:S_RSB}.

Sec.~\ref{sec:level4} was dedicated to the quasi-instantaneous injection of soft photon backgrounds, where we considered injection redshifts at $z_{\rm inj} = 3000, \, 1000, \, 500, \, 100$ and spectral indices $\gamma = 3.0$ (free-free like), $3.3$, and $3.6$ (synchrotron like). Two major conclusions were reached in this section. First, if injection takes place in the pre-recombination plasma ($z \gtrsim 1000$), the low frequency part of the spectrum where $\tau_{\rm ff}(x,z) \gtrsim 1$ is quickly absorbed. The photons that survive this initial absorption into the post-recombination universe exhibit little free-free heating at the redshifts relevant to $21$cm observations, leading to a large boost in the predicted $\Delta T_{\rm b}$, which comes purely from the relative increase in $T_{\rm R}$ compared to the vanilla $\Lambda$CDM scenario.

However, if injections take place after the universe is at least partially recombined, we are met with two important phases of soft photon heating. The first phase again takes place near the injection redshift, where a truncation of the spectrum immediately occurs, but is less severe than in the case of pre-recombination injection due to the reduced $X_{\rm e}$ fraction. The second phase of soft photon heating takes place right around cosmic dawn. When the first sources start to reionize the universe, $X_{\rm e}$ begins to increase rapidly, and a large fraction of the remaining soft photon spectrum is absorbed by the gas. The relative strength of this effect depends on the spectral index of the injected photons. For $\gamma = 3.0$, the effect is largely negligible. Anything steeper than this, however, will introduce important changes to observable quantities such as $\Delta T_{\rm b}$.

We also considered the more physical scenario in which a soft photon background is built up gradually over time. In the decaying dark matter scenario, we studied a fiducial decay rate of $\Gamma = 2\times 10^{-14} \, {\rm s}^{-1}$ for the particle. This causes soft photons to be injected into the background at a continuous rate until a redshift of approximately $z_{\rm dec} \simeq 420$, after which the injection stops. The results obtained in this scenario were qualitatively very similar to what one would expect from a quasi-instantaneous injection around $z_{\rm inj} = z_{\rm dec}$. We additionally considered a model of soft photon emission from superconducting cosmic strings. This model produces soft photons at all redshifts, providing a consistently strong source of heating as evidenced by the upper left panel of Fig.~\ref{fig:CS-overview}. As expected, models of continuous heating such as these do not possess a characteristic initial period of soft photon heating (unlike the quasi-instantaneous injection cases). Importantly, however, is that the consistent production of radiation during the dark ages serves to fill up a soft photon reservoir. The energy stored in this reservoir is then released near cosmic dawn as the first sources begin turning on due to the rapid increase of $X_{\rm e}$.

Soft photon heating also has important implications for 21cm power spectrum calculations as can be seen in Fig. \ref{fig:Pk_21cmfast}. For a homogeneous and isotropic soft photon injection, this heating reduces the temperature contrast between the hot sources and the background. Similar to the global case, this causes a net reduction in the amplitude of the fluctuations, making the signal more difficult to detect. However, the deposition of energy from soft photon heating is expected to occur near to the source. Therefore, sources of anisotropic soft photon injection may lead to new contributions to the power spectrum that one could expect from, e.g., radio loud galaxies. We plan to explore this avenue in a future study.

Recently, there have been papers which aim to provide constraints on the presence of extra radio backgrounds during reionization and cosmic dawn \citep{Mondal2020, HERA2022, Bevins2022}. To the best of our knowledge, these works have not included the effects of soft photon heating, or more generally a correlation between heating processes and significant radio emission, which we have proven here can cause leading order corrections to the differential brightness temperature, both on the global signal and its fluctuations. Unfortunately, soft photon heating serves to severely dampen the amplitude of $\Delta T_{\rm b}$ during this epoch. We expect that once this effect is taken into account, the derived constraints on radio backgrounds will be significantly relaxed.

Soft photon heating is a universal effect that takes place even in a scenario where the CMB is the sole source of radio photons. While for standard scenarios the effect is likely negligible, much uncertainty surrounds our understanding of which low-frequency backgrounds may permeate the universe throughout its thermal history. In this work, we elucidated the proper interplay between these soft photon backgrounds, the matter temperature, and the key observables relevant to $21$cm cosmology. Considering the potential importance of soft photon heating, this effect should be implemented into future studies on exotic energy injections, as well as in forecasting and parameter inference for current and future experiments.

\section*{Acknowledgements}
%
BC would like to acknowledge support from an NSERC postdoctoral fellowship.
SA is supported by an ARCO fellowship.
JC was supported by the ERC Consolidator Grant {\it CMBSPEC} (No.~725456) and by the Royal Society as a Royal Society University Research Fellow at the University of Manchester, UK (No.~URF/R/191023).

\section{Data availability}
The data underlying in this article are available in this article and can further be made available on request.

{\small
\vspace{-3mm}
\bibliographystyle{mn2e}
\bibliography{Lit}
}

\appendix

\section{Formalising the photon evolution equation}
\label{sec:levelA1}
The evolution of the photon occupation number depends primarily on three physical processes (and a potential source term) \citep{Chluba2011therm}
%
\begin{align}
    \frac{\id n_{\rm x}}{\id \tau} = \left.\frac{\id n_{\rm x}}{\id \tau}\right|_{\rm C} +\left.\frac{\id n_{\rm x}}{\id \tau}\right|_{\rm DC} +\left.\frac{\id n_{\rm x}}{\id \tau}\right|_{\rm BR} + \left.\frac{\id n_{\rm x}}{\id \tau}\right|_{\rm Src}.
\end{align}
%
A detailed discussion of the modelling of the source term for quasi-instantaneous injections is given in Section \ref{sec:level4}, so we take the opportunity now to briefly describe the effects of the other three standard processes.

\subsection{Compton interactions}
The first term encodes the Compton process, and describes the effects of energy redistribution, in contrast to double Compton and Bremsstrahlung, which are mainly responsible for the production and absorption of photons. The Compton term is treated by the Kompaneets equation \citep{Kompa56}
%
\begin{align}
    \left.\frac{\id n_{\rm x}}{\id \tau}\right|_{\rm C} = \frac{\theta_{\rm e}}{x^2} \frac{\partial }{\partial x} x^4 \left[ \frac{\partial n_{\rm x}}{\partial x} + \frac{T_{\rm CMB}}{T_{\rm M}} n_{\rm x}(n_{\rm x}+1)\right].
\end{align}
%
Here,  we neglect relativistic effects as they present sub-percent level corrections at the redshifts considered in this work. We also note that $\theta_{\rm e} = k_{\rm b} T_{\rm M}/m_{\rm e}c^2$ where $m_{\rm e}$ is the electron mass. By decomposing the photon spectrum into a blackbody part and a distortion, $n_{\rm x} = n_{\rm bb} + \Delta n$, we find
%
\begin{align} \label{eq:Kompaneet-dist}
    \left.\frac{\id \Delta n}{\id \tau}\right|_{\rm C} = (\theta_{\rm e} - \theta_{\gamma}) \frac{x {\rm e}^x}{({\rm e}^x-1)^2}\left[ x \frac{{\rm e}^x+1}{{\rm e}^x-1}-4\right] + \mathcal{O}(\theta_{\rm e}\Delta n),
\end{align}
%
where $\theta_{\gamma} = \theta_{\rm e} T_{\rm CMB}/T_{\rm M}$ (recall also that $\id n_{\rm bb}/\id \tau = 0$). Here we neglect terms of order $\theta_{\rm e}\Delta n$ and higher not because they are small ($\Delta n$ is not necessarily smaller than $n_{\rm bb}$ for soft photon injections at small $x$), but because at late times, the scattering $y$-parameter, leading to redistribution of photons in energy, becomes unimportant \citep[e.g., Fig. 1 of][]{Chluba2015GreensII}. In addition,
at the low $x$ values relevant to our analysis, the timescale for free-free absorption is much more rapid than the Compton process. Thus, even though $\Delta n$ becomes large at low frequencies, these photons are absorbed by the medium before they have sufficient time to redistribute their energy across the spectrum through Compton interactions. The lowest order term in this expression corresponds to the second source term given in Eq.~\eqref{eq:source-term}.

\subsection{Double Compton}
At early times ($z \gtrsim 10^5$), double Compton scattering plays a crucial role in fully thermalizing distortions that may be induced in the primordial plasma. At late times, however, this process of $1\rightarrow 2$ photon conversions becomes incredibly inefficient relative to that of free-free absorption and emission. As a result, we neglect contributions from this channel in our analysis, but the interested reader is referred to Sec. 2 of \citet{Chluba2011therm}. An in depth discussion of the double Compton process can be furthermore found in \citep{Ravenni2020DC}.

\subsection{Free-free effects}
Bremsstrahlung effects can be included into the photon evolution equation following the procedure laid out by many authors \citep{Rybicki1979, Lightman1981, Danese1982}, taking the form
%
\begin{align}
    \left.\frac{\id n_{\rm x}}{\id \tau}\right|_{\rm BR} = \frac{1}{x^3}\left[ 1 - n_{\rm x}({\rm e}^{x_{\rm e}}-1) \right] \times K_{\rm BR}(x,\theta_{\rm e}).
\end{align}
%
In this expression, $K_{\rm BR}$ is the Bremsstralung emission coefficient, given by \citep{Burigana1991, Hu1993}
%
\begin{align} \label{eq:BR-coll}
    K_{\rm BR}(x,\theta_{\rm e}) = \frac{\alpha \lambda_{\rm e}^3}{2\pi \sqrt{6\pi}} \frac{\theta_{\rm e}^{-7/2} {\rm e}^{-x_{\rm e}}}{(T_{\rm CMB}/T_{\rm M})^3} \sum_{\rm i} Z_{\rm i}^2 N_{\rm i} g_{\rm ff}(z_{\rm i},x,\theta_{\rm e}).
\end{align}
%
Here, $\alpha \simeq 1/137$ the fine structure constant, $\lambda_{\rm e} = h/m_{\rm e} c$ the Compton wavelength of the electron, as well as the charge ($Z_{\rm i}$) and number density ($N_{\rm i}$) of a particular atomic species. The Gaunt factor ($g_{\rm ff}$) in the absence of Helium can be approximated as\footnote{Note there is a typo in Eq.~(31) of \citet{Chluba2015GreensII}, although the implementation was not affected.} \citep{Draine2011Book}
%
\begin{align}
    g_{\rm ff}(x_{\rm e},\theta_{\rm e}) \approx 1 + {\rm ln}\left[1+ {\rm exp} \left( \frac{\sqrt{3}}{\pi} \left[ {\rm ln} \left( \frac{2.25}{Z_{\rm H} x_{\rm e}}\right)+ \frac{{\rm ln}\, \theta_{\rm e}}{2}\right]+1.425\right)\right]. \nonumber
\end{align}
%
Although a more accurate representation can be achieved with {\tt BRpack} \citep{Chluba2020BRpack}, this approximation suffices for our purposes. Inserting all of this into Eq.~\eqref{eq:BR-coll}, one finds
%
\begin{align} \label{eq:ff-starting}
    \left.\frac{\id n_{\rm x}}{\id \tau}\right|_{\rm BR} = \frac{{\rm e}^{-x_{\rm e}}}{x_{\rm e}^3}\left[ 1 - n_{\rm x}({\rm e}^{x_{\rm e}}-1) \right]  \Lambda_{\rm BR},
\end{align}
%
where we have now defined the Bremsstrahlung emissivity coefficient, $\Lambda_{\rm BR}$ as
%
\begin{align}
    \Lambda_{\rm BR}(\tau, x_{\rm e}) = \frac{\alpha \lambda_{\rm e}^3}{2\pi \sqrt{6\pi}} \theta_{\rm e}^{-7/2} N_{\rm p} \,g_{\rm ff}(x_{\rm e},\theta_{\rm e}),
\end{align}
%
with $N_{\rm p}$ being the proton number density. This matches with the starting point in our computation, namely Eq.~\eqref{eq:starting-point}.

\section{Matter temperature evolution}
\label{sec:levelB1}
Using the first law of thermodynamics, one can study the evolution of the gas temperature in the presence of a photon background. The general expression is given by 
%
\begin{align} \label{eq:temp-evo-useful}
    \frac{\id T_{\rm M}}{\id z} = \frac{2 T_{\rm M}}{1+z} + \frac{1}{\chi \, a^3}\frac{\id Q_{\rm M}}{\id z} - \frac{1}{\chi \, a^4} \sum_i  \frac{\id (a^4 \Delta \rho_{\gamma,i})}{\id z},
\end{align}
%
where $Q_{\rm M}$ is the direct (comoving) energy density injection into the gas, $\Delta \rho_{\gamma}$ is the energy density in the various constituents of a spectral distortion ($\Delta \rho_{\gamma} = \sum_i \Delta \rho_{\gamma, i}$, with $\rho_{\gamma} = \rho_{\rm CMB} + \Delta \rho_{\gamma}$), and 
%
\begin{align}
    \chi = \frac{3}{2} k_{\rm b} N_{\rm H}(1+X_{\rm e} + f_{\rm He})
\end{align}
%
is related to the heat capacity of the medium. We consider no direct energy injections to the gas ($\!\id Q_{\rm M}/\id z = 0)$, and discuss contributions from both Compton scattering and free-free.

\subsection{Compton heating}
As discussed above, Kompaneets equation governs to the evolution of the photon occupation number in the presence of Compton scatterings between the background and the matter sector. Neglecting again the relativistic corrections, Eq.~\eqref{eq:Kompaneet-dist} takes a useful form
%
\begin{align}
    \left.\frac{\id \Delta n}{\id \tau} \right|_{\rm C} \approx \frac{(\theta_{\rm e} - \theta_{\gamma})}{x^2} \frac{\partial}{\partial x} x^4 \left( \frac{\partial n_{\rm bb}}{\partial x}\right),
\end{align}
%
where we have made use of the identity $\partial n_{\rm bb}/\partial x = - n_{\rm bb}(n_{\rm bb} + 1)$. For an isotropic photon distribution, the (physical) energy density is
%
\begin{align}
    \rho = \frac{(k_{\rm b} T_{\rm CMB})^4}{\pi^2 \hbar^3 c^3} \int_0^{\infty} \id x \, x^3 \, n(x).
\end{align}
Thus, the photon energy injection rate is given by 
%
\begin{align}
    \frac{1}{a^4} \frac{\id (a^4\Delta \rho_{\rm C})}{\id \tau} &= \frac{(k_{\rm b} T_{\rm CMB})^4}{\pi^2 \hbar^3 c^3} \int_0^{\infty} \id x \, x^3 \, \frac{\id \Delta n_{\rm C}}{\id \tau},\nonumber \\
    &=4 \rho_{\rm CMB} (\theta_{\rm e} - \theta_{\gamma}),
\end{align}
%
where $\rho_{\rm CMB}$ is the energy density of the CMB blackbody.
The conversion to redshift is given by $\id \tau = -[\sigma_{\rm T} N_{\rm e} c/H(1+z)] \id z$, which upon substitution into Eq.~\eqref{eq:temp-evo-useful} yields the well known Compton heating term, the second term in Eq.~\eqref{eq:T_M}.

\subsection{Free-free heating}
Free-free heating plays a negligible role without the presence of an enhancement to the low frequency ($x \lesssim 1)$ part of the photon spectrum. The free-free part of the injection can be seen by insertion of $n_{\rm x} = n_{\rm bb} + \Delta n$ into Eq.~\eqref{eq:ff-starting}, which yields the evolution equation and source terms found in Sec. \ref{sec:level3}. This procedure provides
%
\begin{subequations}
\begin{align}
    \frac{\id \Delta n_{\rm ff,bb}}{\id \tau} &= \frac{\Lambda_{\rm BR} \left(1- {\rm e}^{-x_{\rm e}}\right)}{x_{\rm e}^3} \left[\frac{1}{{\rm e}^{x_{\rm e}}-1} - \frac{1}{{\rm e}^{x}-1}\right],\\
    \frac{\id \Delta n_{\rm ff,dist}}{\id \tau} &= -\frac{\Lambda_{\rm BR} (1 - {\rm e}^{-x_{\rm e}})}{x_{\rm e}^3} \Delta n.
\end{align}
\end{subequations}
%
The first expression carries information about free-free absorption and emission off the blackbody, while the second does the same in the presence of a distortion. Computing the energy density present in this expression yields 
%
\begin{align} 
    \frac{1}{a^4} &\frac{\id (a^4 \Delta \rho_{\rm ff})}{\id z} =
    \frac{(k_{\rm b} T_{\rm CMB})^4}{\pi^2 \hbar^3 c^3} \left( \frac{T_{\rm M}}{T_{\rm CMB}}\right)^3 \frac{\sigma_{\rm T} N_{\rm e} c}{H(1+z)} \\
    &\times \int_0^{\infty} \id x \, \Lambda_{\rm BR}(\tau, x_{\rm e}) \left( 1 - {\rm e}^{-x_{\rm e}}\right) \left[ \Delta n(x) + \frac{1}{{\rm e}^{x}-1} - \frac{1}{{\rm e}^{x_{\rm e}}-1} \right]. 
\nonumber
\end{align}
%
Clearly, the presence of large enhancements to $\Delta n$(x) (such is the case for extra radio backgrounds) can significantly boost the heating rate, leading to a rise in the overall matter temperature.

\vspace{-5mm}

\end{document}